\documentclass[pdflatex,sn-mathphys-ay, iicol]{sn-jnl}

\usepackage{graphicx}%
\usepackage{multirow}%
\usepackage{amsmath,amssymb,amsfonts}%
\usepackage{amsthm}%
\usepackage{mathrsfs}%
\usepackage[title]{appendix}%
\usepackage{xcolor}%
\usepackage{textcomp}%
\usepackage{manyfoot}%
\usepackage{booktabs}%
\usepackage{algorithm}%

\usepackage{algpseudocode}%

\usepackage{listings}%

\theoremstyle{thmstyleone}%
\theoremstyle{thmstyletwo}%

\theoremstyle{thmstylethree}%

\begin{document}

\title[A Scoping Review of Physics Informed Machine Learning for Wave Propagation Modeling in Seismology]{\textbf{A Scoping Review of Physics Informed Machine Learning for Wave Propagation Modeling in Seismology}}









\author[1]{\fnm{O.} \sur{Rincón-Cardeño}}\email{orincon@eafit.edu.co}

\author[1]{\fnm{G.} \sur{Pérez-Bernal}}\email{gperezb1@eafit.edu.co}

\author[2]{\fnm{S.} \sur{Montoya-Noguera}}\email{smontoyan@eafit.edu.co}

\author*[1]{\fnm{N.} \sur{Guarín-Zapata}}\email{nguarinz@eafit.edu.co}

\affil*[1]{\orgdiv{Mathematical Applications in Science and Engineering Research Group, School of Applied Sciences and Engineering}, \orgname{Universidad EAFIT}, \orgaddress{\city{Medellín}, \country{Colombia}}}

\affil[2]{\orgdiv{Nature and Cities Research Group, School of Applied Sciences and Engineering}, \orgname{Universidad EAFIT}, \orgaddress{\city{Medellín}, \country{Colombia}}}

\abstract{Standard numerical methods accurately simulate seismic waves but are computationally expensive, particularly for inverse problems. Some machine-learning-based alternatives have emerged, promising to reduce cost while preserving physical accuracy. This scoping review, drawing on OpenAlex and Scopus, maps physics-informed machine learning applications to seismic wave propagation based on partial differential equations, classifying selected studies by problem type (forward or inverse) and learning strategy to uncover trends, patterns, and gaps. Results show that these methods have been applied to both forward modeling and inversion, often matching standard numerical accuracy at lower cost. We identified three mechanisms for incorporating physical knowledge into models: observational, inductive, and learning bias. We implemented a PyTorch replication of the original PINN framework to test methodological reproducibility of a representative method, obtaining results consistent with and, in most cases, more accurate than those originally reported. Based on the reviewed literature, we identified limitations in benchmarking consistency, training cost, and scalability to three-dimensional and experimentally validated problems. We conclude that while standard numerical methods remain foundational, physics-informed machine learning serves as a complementary tool valuable for inverse problems and surrogate modeling, with future work needing consistent benchmarking, hybrid formulations, and validation under realistic geophysical conditions. The continued development of these methods, particularly through hybrid formulations and reproducible benchmarking, may broaden their role in seismological workflows, although realizing this potential will need addressing current limitations in scalability.}
 
\keywords{wave propagation, seismology, physics-informed neural networks, full waveform inversion, surrogate modeling, partial differential equations, numerical methods.\\
\vspace{-0.2cm}
}

\maketitle

\section*{Introduction}\label{sec1}
 
Wave propagation is a physical phenomenon described by partial differential equations (PDEs), which are widely used due to their applicability in fields such as medicine and seismology. However, analytical solutions are available in many practical situations, and therefore numerical methods are generally required to obtain approximate solutions \citep{seriani_numerical_2020}. In the study of wave propagation, numerous methods have been developed to address the associated physical and numerical challenges. Among the most widely used standard numerical approaches are the finite difference method, the finite element method, and the boundary element method \citep{igel_computational_2017,virieux_review_2011}. These methods discretize the spatial domain to obtain approximate solutions to the governing differential equations. 

To reduce the relative error with respect to the analytical solutions, one improves the numerical solution by refining the mesh. This simultaneously increases the computational cost. Let us consider the solution to the Helmholtz equation:

\begin{equation}
\begin{cases}
\nabla^2 f + (5\pi)^2 f = 0, & \text{in } \Omega\, , \\[6pt]
f = 0, & \text{on } \partial\Omega\, .
\end{cases}
\label{eq:helmholtz2D}
\end{equation}

\vspace{0.2cm}

\indent The corresponding analytical solution is given by
\[
f(x,y) = \sin(5\pi x)\sin(5\pi y)\, .
\]

\vspace{0.2cm}

\begin{figure*}[ht!]
    \centering
    \includegraphics[width=6.2 in]{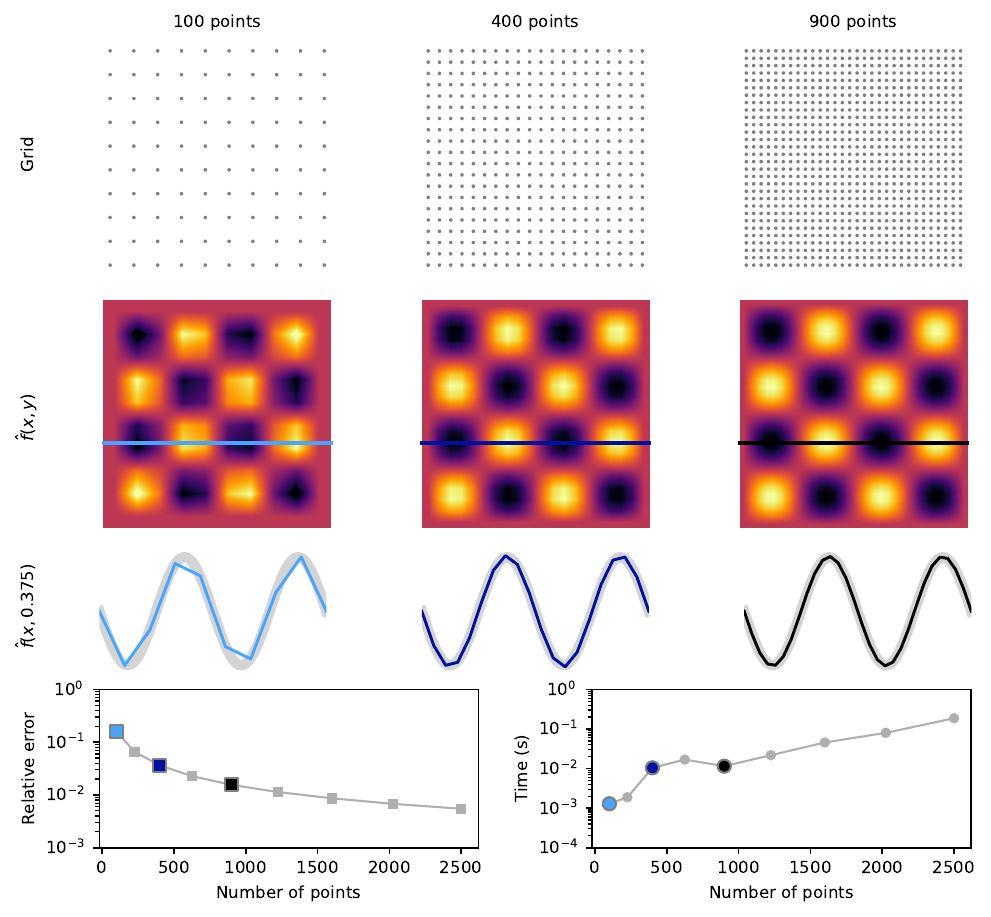}
\caption{Convergence of the numerical solution of the Helmholtz equation in 2D. (Top row) Point distributions used in the simulations, corresponding to $N = 10 × 10$, $20 × 20$, and $30 × 30$.(Second row) Contour maps of the approximate solution $\hat{f}(x,y)$ for each solution. (Third row) One-dimensional cross-sections of $\hat{f}(x,0.375)$ comparing the spatial variation of the approximate solution (light blue, dark blue, and black lines) with the analytical solution (gray line). (Bottom row) Convergence plots showing the relative error and computation time as a function of the total number of points.}
    \label{fig:helmholtz2D_convergence}
\end{figure*}

\indent We present numerical approximations, \(\hat{f}(x, y)\), to the equation using the finite difference method in Figure \autoref{fig:helmholtz2D_convergence}. We can see how the mesh refinement improves the numerical approximation, reduces the relative error compared to the analytical solution, and simultaneously increases computation time. 

In computational modeling, a primary objective is to ensure that solution methods are efficient and exhibit a sufficiently low error to capture the inherent physical details of the system required by the problem of interest. However, standard numerical methods often face computational challenges when dealing with complex problems, particularly in unbounded domains and inverse problems \citep{gosselin_review_2022}. Consequently, the computational demands associated with many common models have motivated the development of innovative solution strategies.

\begin{figure*}
\centering
    \includegraphics[width=6.2 in]{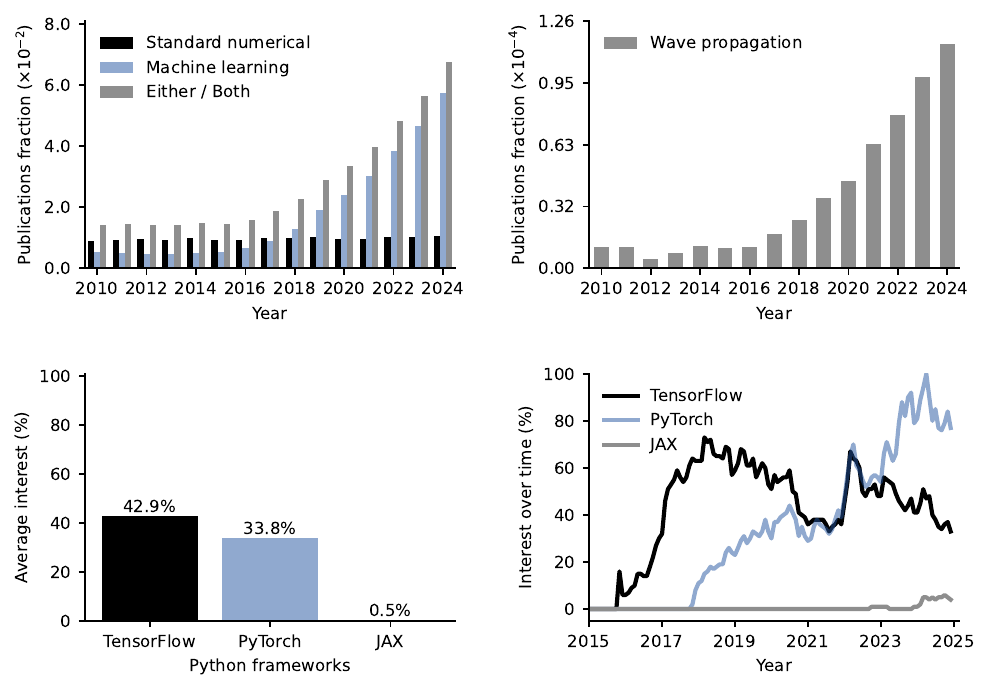}
\caption{Growth of the literature and trends in machine learning Python frameworks.
The upper-left panel shows the fraction of publications related to standard numerical methods and machine learning, while the upper-right panel presents those specifically associated with wave propagation modeling.
The lower-left panel shows the average relative interest in the main machine learning frameworks (TensorFlow, PyTorch, and JAX), computed as the average of their normalized Google search values over the considered time period.
The lower-right panel presents the temporal evolution of this interest, measured as Google searches relative to the maximum observed value (2015--2025).
Relative publication values were computed with respect to the total number of publications indexed in Scopus over the same period (2010--2024).}
    \label{fig:publications_trends}
\end{figure*}
 
In this context, machine learning-based implementations may improve the computational efficiency of certain applications, such as seismic wave propagation modeling, which has traditionally been addressed using standard numerical methods. In contrast to machine learning, whose rapid development is relatively recent, numerical methods have a well-established history spanning several decades, closely tied to advances in computational power, the automation of complex tasks, and a solid mathematical foundation supported by rigorous theoretical results in various contexts \citep{rude_research_2018,burden_numerical_2016}. 

These methods have reached a high level of maturity, with successful applications across numerous fields of science and engineering. Nevertheless, the rapid evolution of machine learning opens new opportunities to tackle problems that have long been addressed using standard numerical approaches. \autoref{fig:publications_trends} shows the relative number of publications, as a percentage of the total number of indexed publications, that link machine learning methods and standard numerical methods to the modeling of partial differential equations, with a particular focus on wave propagation modeling, from 2010 to 2024. Among these works, Physics-Informed Neural Networks (PINNs) have garnered significant attention in recent years, being cited in over 10,000 publications \citep{raissi_physics-informed_2019}. The figure also illustrates the growing interest in three of the most widely used Python frameworks since 2015: TensorFlow, PyTorch, and JAX.

These data reflect an increasing scientific interest in machine learning methods, as well as the growing integration of standard numerical methods with machine learning, with a notable rise in publication rates since 2016. This trend has been driven by advances in hardware, such as graphics processing units (GPUs), a substantial increase in data availability, and the development of open-source computational frameworks. These factors have significantly enhanced the accessibility and applicability of machine learning methods across various disciplines. Notably, the emergence of widely adopted Python frameworks, including TensorFlow, PyTorch, and JAX, has further facilitated the implementation of these methods, making them more accessible to researchers and promoting their adoption in multidisciplinary contexts.

Machine learning has emerged as a computational tool for modeling systems governed by physical laws and described by partial differential equations \citep{cuomo_scientific_2022}, in which the function to be approximated represents the variable of interest. Machine learning methods offer significant advantages in terms of evaluation speed and adaptability, as they can be adjusted to different medium configurations, boundary conditions, or data types without the need to fully reformulate the underlying physical model. Moreover, they are efficient in contexts where equation-based models are computationally expensive or difficult to construct, positioning them as tools with high potential. Nevertheless, they also present important limitations, such as limited physical interpretability \citep{zhang_survey_2020}.

In particular, neural networks are capable of learning relationships between input features and output targets from examples. Two important applications are classification and function approximation. A version of the universal approximation theorem, as established in \citep{hornik_approximation_1991}, states that a feedforward neural network (FFNN) with a single hidden layer and a finite number of neurons can arbitrarily approximate any continuous function defined on a compact subset of the input space, provided that the activation function is non-constant, bounded, and continuous. To illustrate this concept, let's consider the following function:
\[
f(x) = x^3 + x^2 - x - 1\, .
\]

\begin{figure*}[ht!]
    \centering
    \includegraphics[width=5.5 in]{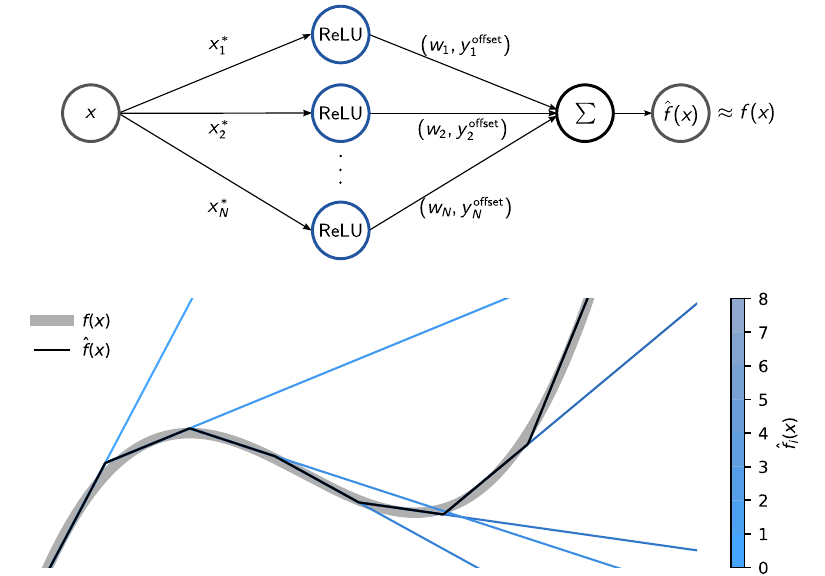}
    \caption{Approximation of an objective function using a hidden single-layer neural network with varying numbers of neurons. The gray curve represents the original function, while the blue curves show the approximations obtained by gradually increasing the number of neurons. The black curve is the final approximation with the total number of neurons.}
    \label{fig:universal_approximation_demo}
\end{figure*}

To approximate this function, we divide the domain into \(N\) segments of equal length. In each segment, a contribution is defined based on the ReLU (Rectified Linear Unit) activation function, so that the segment is represented as
\[
g_i(x) = w_i \, \text{ReLU}(x - x_i^*) + y^{\text{offset}}_{i}\, ,
\]
where \(x_i^*\) is the start of the segment, \(w_i\) is the weight that determines the slope, and \(y^{\text{offset}}_{i}\) is a vertical offset that ensures continuity with previous segments. The ReLU function ensures that the contribution of each segment starts in its corresponding region and is defined as
\[
\text{ReLU}(x - x_i^*) = \max(0, x - x_i^*)\, ,
\]
generating a piecewise linear function in each segment. \autoref{fig:universal_approximation_demo} illustrates this process, showing how each segment activated by ReLU contributes to the reconstruction of the objective function. The complete approximation is obtained after all segments have been processed
\[
f(x) \approx \hat{f}_{i}(x) = \sum_{i=1}^{N} g_i(x)\, .
\]
\indent Each segment adds a ReLU-activated straight line that contributes locally to the shape of the function. Combining all the segments produces a piecewise approximation that captures the structure of \(f(x)\).

As the number of segments increases, the approximation gains accuracy and becomes capable of representing more complex functions. This representational capacity is known as the network's expressiveness. This procedure exemplifies the ability of neural networks to approximate functions. However, we must also consider that there is no guarantee that the training algorithm will be able to learn the desired function. The optimizer might not find the appropriate parameter values or might choose an incorrect function due to overfitting. Therefore, as noted by \citep{goodfellow_deep_2016}, although a single-layer FFNN can approximate any function, such a layer might require an impractical number of neurons and still not guarantee adequate learning and generalization.

A method that seeks to preserve the advantages of machine learning while explicitly incorporating physical knowledge is PINNs \citep{raissi_physics-informed_2019}. Altought it have been applied to a wide range of problems, challenges remain when comparing their performance with standard numerical methods, largely due to differences in their formulation. PINNs also enable the solution of inverse problems \citep{raissi_hidden_2020,haghighat_physics-informed_2021,hao_physics-informed_2023}, in which physical parameters of a model are inferred from experimental or observational data. This capability is particularly valuable in seismology, where subsurface characterization relies on methods capable of effectively combining observational data with physical models of wave propagation. Therefore, PINNs emerge as a promising alternative for the efficient estimation of model parameters.

Several reviews have examined the use of machine learning across different areas of science and engineering (e.g. \citep{vadyala_review_2022,deng_physics-informed_2023,lino_current_2023}), including applications to seismic inversion problems \citep{jingbo_research_2023}. Nevertheless, given the rapid expansion of the field, uncertainty remains regarding which techniques have have demonstrated advantages over standard methods, and under what conditions \citep{mcgreivy_weak_2024,grossmann_can_2024}.

In principle, machine learning approaches have the potential to learn surrogate models, that is, approximate representations of physical systems allowing predictions to be obtained without directly solving the governing equations. Such surrogates can significantly reduce evaluation time, a critical advantage in computational seismology, where computational demands are considerable. However, this advantage is not always analyzed in the existing literature. Therefore, an examination of these approaches is necessary to provide a current overview of the state of the art, to identify their current applications, and open challenges that may guide future investigations. We adopted a scoping review approach to map physics-informed machine learning in wave propagation modeling \citep{arksey_scoping_2005}. Particular attention is given to seismic inversion, where they are used to infer subsurface properties from data. This review categorizes existing methodological approaches, and discusses the advantages and limitations of machine learning relative to standard numerical methods.

\section*{Methods}\label{sec:methods}

This review was conducted following the PRISMA Extension for Scoping Reviews (PRISMA-ScR) guidelines \citep{tricco_prisma_2018}, and the completed PRISMA-ScR checklist is provided as supplementary material. As outlined by \cite{arksey_scoping_2005}, a scoping review aims to ``map rapidly the key concepts underpinning a research area and the main sources and types of evidence available'' \citep{arksey_scoping_2005}. In line with this objective, the review process was structured into five sequential stages:

\begin{enumerate}
    \item Identifying the research question
    \item Identifying relevant studies
    \item Selecting studies to be included in the review
    \item Charting the data
    \item Collating, summarizing, and reporting the results
\end{enumerate}
 
\begin{figure*} 
    \includegraphics[width=6.5 in]{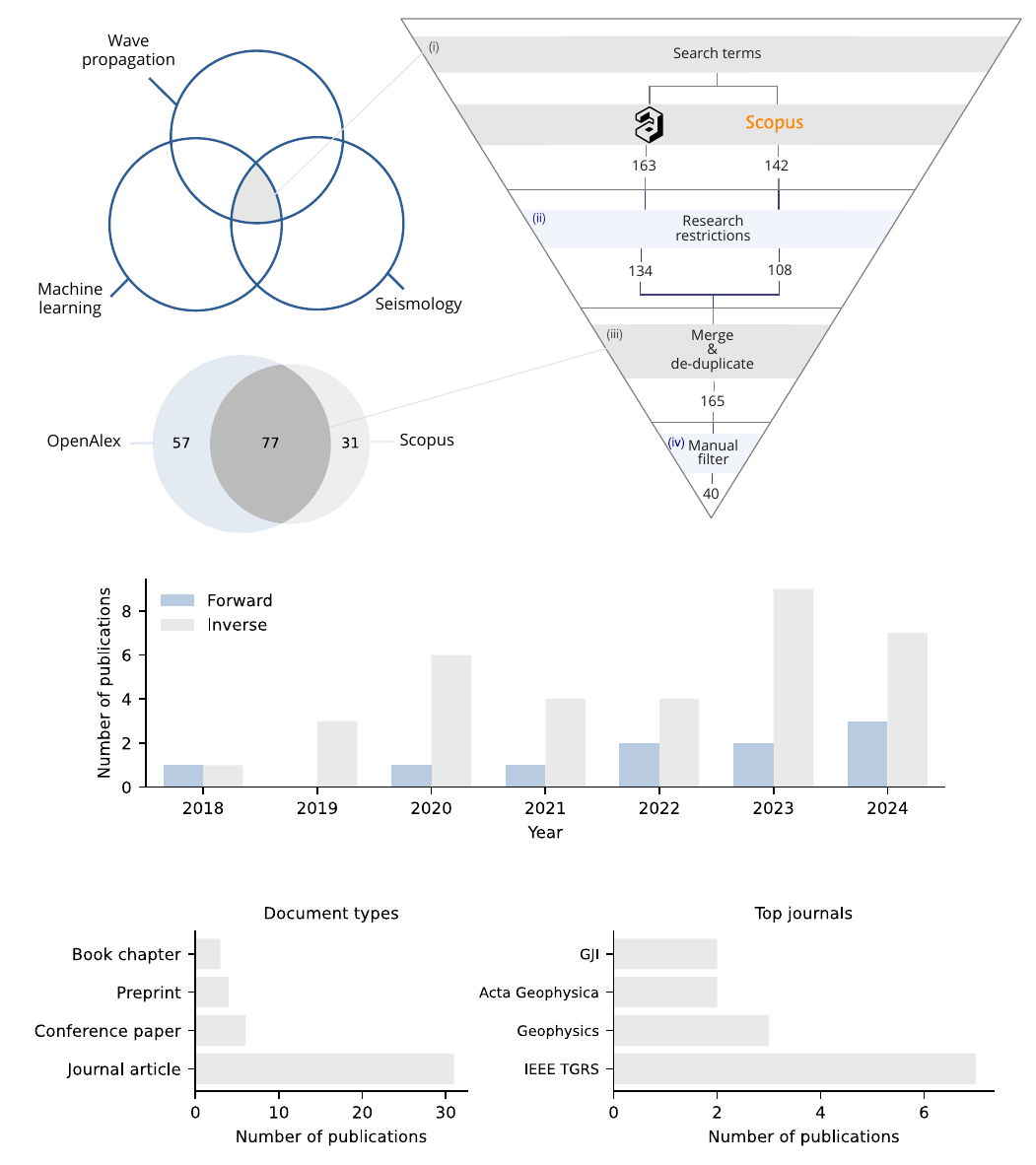}
    \vspace{-0.5cm}
\caption{Overview of the scoping review workflow and descriptive statistics of the selected publications. 
The conceptual Venn diagram (top left) illustrates the intersection of the three main research areas considered in the search strategy: wave propagation, machine learning, and seismology. 
The workflow funnel (top right) summarizes the screening process: (i) search terms were applied in OpenAlex and Scopus, resulting in 163 and 142 records, respectively; 
(ii) research restrictions were applied, retaining 134 and 108 records; 
(iii) records from both databases were merged and duplicates were removed, resulting in 165 unique publications, as shown by the database overlap diagram; 
(iv) a final manual screening based on abstract evaluation identified 40 studies relevant to the research objectives. 
The middle panel presents the temporal distribution of the selected publications from 2018 to 2024, distinguishing between forward modeling and inverse problem formulations. 
The bottom panels summarize the descriptive characteristics of the final dataset, including the distribution by document type (left) and the most frequent journals (right).}
\label{fig:scheme_systematic_review}
\end{figure*}

The workflow of the literature selection process is summarized in \autoref{fig:scheme_systematic_review}. These methodological steps were designed to ensure reproducibility.\\

\noindent\textbf{Stage 1: Identifying the research question.} This stage defines the scope of the review through the formulation of the central research question that guides the study. The review seeks to address the following question:\\

\noindent What physics-informed machine learning methods have been applied to model wave propagation in seismology?\\
 
Although this area has broad applicability, the scope of this review is restricted to seismic wave propagation. The research questions therefore emphasize applications where machine learning methods complement or serve as alternatives to standard numerical approaches to wave equation modeling. This includes cases where synthetic data generated using standard numerical solvers are used to train a machine learning model, leading to a reduction in computational cost during inference. Also, scenarios in which machine learning methods are used as standalone solvers and are reported to offer improvements in efficiency relative to traditional techniques.

The scope of the review also includes studies in which machine learning approaches are applied to inverse problems in seismology, where these methods are particularly promising. This interest is motivated by their ability to consider varying amounts of data, incorporate physical constraints directly into the learning process, and reduce the computational time associated with iterative solution strategies that require repeated evaluations of forward models. 

To support the pursuit of the main research question, the review is further guided by the following sub-questions:

\begin{enumerate}
    \item What physics-informed strategies are employed in machine learning-based modeling of seismic wave propagation?
    \item What hybrid methods combine standard numerical solvers with machine learning techniques?
    \item In the context of inverse problems, what evidence is reported regarding the advantages of machine learning-based methods over standard numerical approaches?
    \item What limitations remain in scaling machine learning methods to realistic seismic applications?
\end{enumerate}

Together, these questions provide a framework for mapping and analyzing the existing literature.\\

\noindent\textbf{Stage 2: Identifying relevant studies.} We developed a search strategy based on the research question defined in Stage~1 and refined using predefined inclusion and exclusion criteria. The search aimed to identify studies applying machine learning techniques to seismic wave modeling, with particular emphasis on approaches that incorporate physical laws through PDEs. We used the following search query:\\

\noindent\texttt{("machine learning" OR "deep learning" OR "neural networks") AND ("seismic" OR "seismology") AND "wave equation" AND (modeling OR modelling OR model OR simulation)}\\

We used OpenAlex and Scopus as databases for the literature search. OpenAlex was employed as the primary data source, while Scopus was used to validate and complement the results obtained from OpenAlex. In both databases, searches were restricted to documents containing the specified terms in their titles or abstracts. The inclusion of OpenAlex as a data source was motivated by its open-access nature, which enables transparency and allows readers to independently access and verify the studies included in this review.\\

\noindent\textbf{Stage 3: Selecting studies to be included in the review.} Study selection was guided by inclusion and exclusion criteria. Eligible studies incorporated physical phenomena modeled by PDEs, such as seismic wave equations, in combination with machine learning techniques. Although most of the approaches are based on neural networks, we adopt the broader term machine learning to encompass a wider range of methods beyond neural networks. Also, studies were included when they reported quantitative or supported qualitative comparisons of the computational efficiency of machine learning approaches relative to standard numerical methods. In addition, studies addressing inverse problems in seismology were considered within scope. 

Studies were excluded if they focused exclusively on forward modeling accuracy without addressing computational efficiency or if they did not include a comparison with standard numerical methods. Studies that compared machine learning approaches solely with other machine learning methods, without reference to standard numerical techniques, were also excluded. Furthermore, studies outside the domain of seismology were not considered. We ensured that all included studies met these criteria through a screening process, beginning with a title and abstract review, followed by a full-text review.\\

\noindent\textbf{Stage 4: Charting the data.} 
A data-charting process was developed to provide a descriptive analysis of the included literature. Each selected publication was independently reviewed by at least two authors to extract information relevant to the research questions. The extracted information was then discussed among the authors to reach an interpretation of the findings. During this process, a set of specific questions associated with the objectives of the review was answered and documented for each study, covering aspects such as publication characteristics, methodological approaches, neural network architectures, governing equations, forward and inverse modeling strategies, scalability, evaluation procedures, limitations, and future research directions. The resulting extracted data and annotations from the reviewed publications are available in the associated dataset \citep{rincon_2026_21054377}.

The resulting charted data were used to identify thematic and methodological trends across the selected studies. The code used to implement the data charting derived from the reviewed literature is available at
\url{https://github.com/oscar-rincon/scoping-ml-wave-seismology}.\\
 
\noindent\textbf{Stage 5: Collating, summarizing, and reporting the results.} We organized the extracted information from the selected studies to address the research questions defined in Stage~1 and to identify the main methodological trends in physics-informed machine learning for seismic wave propagation. First, we characterized the reviewed studies according to the type of problem addressed (forward modeling or inverse problems), the governing physical formulation, and the conventional numerical methods used as reference approaches. This provided the necessary context to compare machine learning-based methods with established computational strategies. 

We grouped the studies according to the mechanism through which physical knowledge was incorporated, following the classification of observational bias, inductive bias, and learning bias. This categorization allowed for the identification of how different approaches integrate physical constraints, data, and model architectures. Finally, we synthesized the results by analyzing recurring advantages, limitations, computational trends, and challenges reported across the literature, with particular attention to scalability, reproducibility, generalization to realistic seismological settings, and the potential role of hybrid approaches combining machine learning with conventional numerical solvers.

\section*{Characteristics of the included literature}

In OpenAlex, the search was limited to documents containing the relevant terms in the title and abstract, yielding 163 records across several document types, including journal articles (120), preprints (32), and book chapters (5). A parallel search in Scopus returned 142 records. These records were filtered to retain journal articles (78), preprints (27), conference papers (25), and book chapters (3), and to exclude non-English publications. After filtering, 134 records from OpenAlex and 108 records from Scopus were retained.

The search results obtained from the OpenAlex and Scopus databases were merged, and duplicate records were removed, resulting in 165 unique publications \citep{rincon_2026_21017942}. The abstracts of these records were manually screened according to the predefined inclusion criteria. Furthermore, the reference lists of the selected studies were examined to identify additional relevant publications. Following this process, 40 studies published between 2018 and 2024 were retained for the final review. The complete bibliographic dataset of the included studies is available in a Zenodo repository \citep{rincon_2026_20834562}.

\begin{figure*}
\centering
    \includegraphics[width=6.2 in]{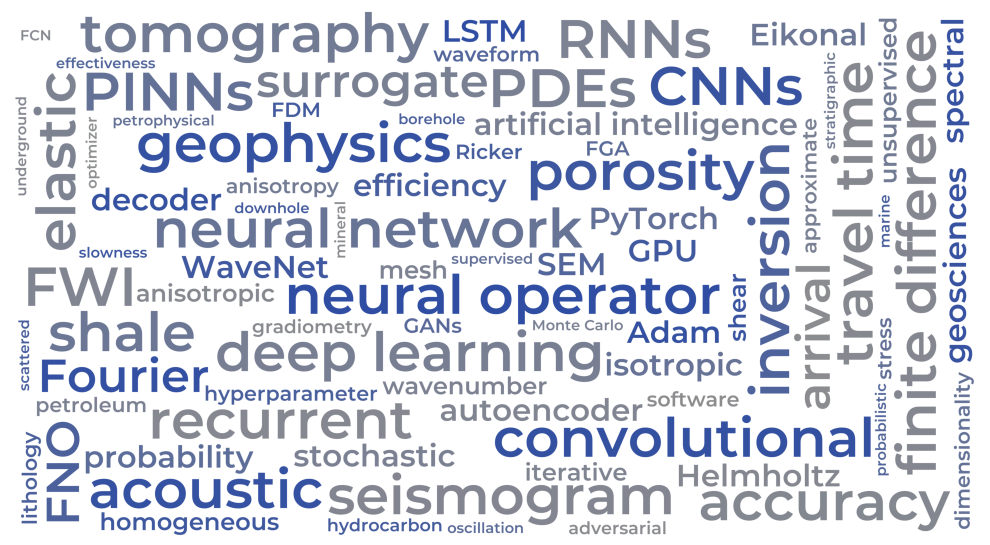}
\caption{Word cloud of the reviewed literature. The figure highlights the most recurrent terms across the 44 selected studies, providing an overview of the dominant topics and methodological trends discussed in the review. Terms were extracted from the full text and processed through lemmatization and part-of-speech filtering (nouns and adjectives only), while word size represents normalized frequency.}
\label{fig:wordcloud}
\end{figure*}

\autoref{fig:wordcloud} presents a word cloud showing the most frequent terms across the reviewed studies. The visualization is based on term frequencies obtained after text preprocessing, including lemmatization, part-of-speech filtering, and the removal of generic and non-informative terms. 

Frequently occurring terms are associated with neural network–based approaches, including their use as surrogate models, as well as with established numerical methods such as finite differences. Terms related to method performance also appear, indicating continued attention to accuracy and computational cost. In addition, full waveform inversion (FWI) appears among the recurring terms, reflecting its relevance in studies addressing inverse problems.
 
\section*{Wave Propagation Modeling: forward and inverse approaches}

In seismology, seismic wave propagation is used to investigate the internal structure of the Earth. Its applications range from energy resource exploration to the identification of active faults and the study of processes associated with plate tectonics \citep{stein_introduction_2009,alsadi_seismic_2017}. It is also used in applied settings such as seismic-while-drilling, where wavefield measurements support the estimation of formation properties \citep{auriol_sensing_2021}. These applications require mathematical descriptions capable of reproducing seismic wave behavior, which motivates the use of mathematical models.

A mathematical model describing wave propagation in a medium aims to represent, through a function, how a system evolves in time and space. Such models are typically based on differential equations, which allow the temporal and spatial evolution of the system to be characterized. In mathematical modeling, two general approaches are commonly employed: the forward problem and the inverse problem. The inverse problem consists of determining the causes from a set of observations \citep{roth_neural_1994,tarantola2005inverse}. For example, inferring the properties of a medium, such as the propagation velocities of different materials, from its response to wave propagation.

This approach is the opposite of the forward problem, in which the effects are computed from known medium properties. Because the inverse problem starts from observed effects and seeks to identify their causes, it typically requires the iterative solution of forward models, making it a computationally demanding process.

\begin{figure*} 
\centering
    \includegraphics[width=6.2 in]{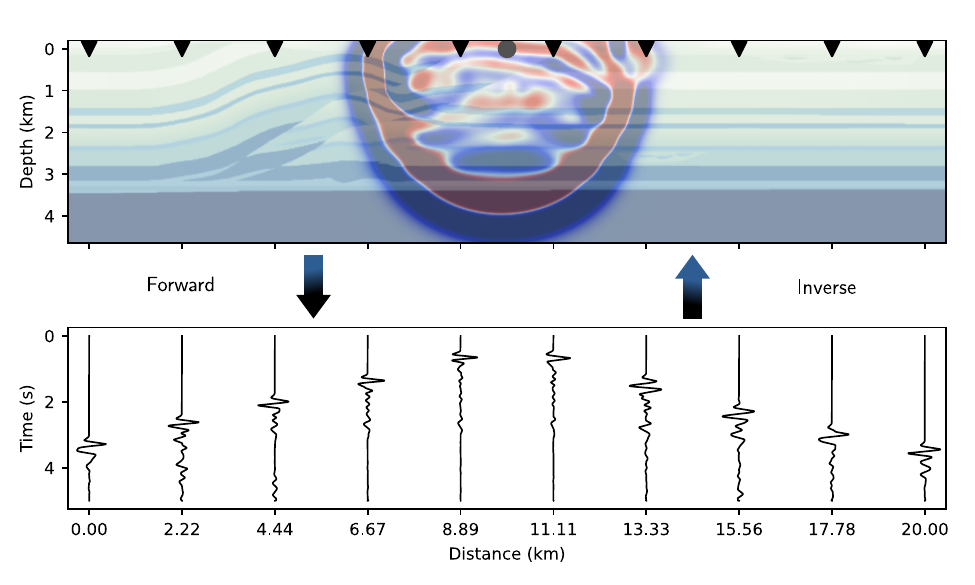}
\caption{Illustrative schematic of a forward and inverse problem in seismology. 
The forward simulation models seismic wave propagation in a given velocity medium, superimposed with the wavefield generated by a point source (blue–red color scale) and recorded by a surface geophone array (black triangles). 
The inverse problem uses the recorded data (seismograms shown at the bottom) to estimate subsurface properties.}
\label{fig:inverse_geophysics}
\end{figure*}

\autoref{fig:inverse_geophysics} schematically illustrates the relationship between forward and inverse problems in seismology: while the former simulates seismic wave propagation through a known velocity model, the latter uses recorded data to infer subsurface velocities. In this example, the Overthrust velocity model \citep{aminzadeh_segeaeg_1994} is used as a reference.

The choice of governing equation is associated with the fidelity and computational cost of both forward and inverse seismic modeling. The scalar acoustic wave equation is widely used to study seismic wave propagation. For example, it has been adopted in physics-based deep learning formulations for FWI \citep{ren_physics-based_2020}. Moreover, under the assumption of time-harmonic solutions, the acoustic wave equation can be transformed into the Helmholtz equation, which is frequently used in seismology \citep{song_high-frequency_2022,song_simulating_2023}. The Helmholtz equation describes the spatial distribution of the wavefield at a given frequency. Elastic wave equations are also employed in multiparameter seismic inversion, where parameter cross-talk remains a well-known challenge \citep{dhara_elastic-adjointnet_2022}. The need to solve these equations efficiently under different conditions has motivated the development of advanced numerical methods and, more recently, the integration of machine learning techniques into seismic modeling workflows.

\section*{From standard numerical methods to physics-informed machine learning}

Over the past decades, several numerical methods have been developed to describe physical systems governed by partial differential equations (PDEs), such as the wave equation. Among them, the finite-difference method is one of the most widely used approaches for wave propagation problems. Its broad use is mainly due to its conceptual simplicity and ease of implementation. A comprehensive review of finite-difference methods for wave propagation is provided by \citep{moczo_finite-difference_2007}. In this approach, partial derivatives are approximated by discrete operators based on differences between neighboring grid points. The finite-difference method is well suited for problems with simple geometries. In contrast, methods such as the finite element method provide greater flexibility in mesh design, which facilitates the treatment of complex geometries.

In numerical simulations of seismic wave propagation, boundary conditions are relevant because the physical subsurface is unbounded, whereas numerical models are solved on finite computational domains. Without appropriate boundary treatment, artificial reflections generated at the domain edges can affect the simulated wavefield and the interpretation of results. For this reason, conventional numerical methods commonly rely on domain truncation techniques such as absorbing boundary conditions (ABCs) or perfectly matched layers (PMLs) to approximate open media and reduce boundary reflections. In the context of PINNs, \cite{zhang_seismic_2023} showed that seismic inversion can be performed effectively with PINNs without explicitly imposing absorbing boundary conditions.

Several open-source implementations are available for applying numerical methods to the solution of the wave equation. For example, SPECFEM, which is specialized in seismic wave propagation, is widely used for simulations in complex geological structures using spectral elements, with implementations primarily written in Fortran \citep{komatitsch_specfemspecfem2d_2023,komatitsch_specfemspecfem3d_2024,hateley_deep_2020}. DEVITO and SEISMIC\_CPML employ finite-difference methods for wavefield modeling \citep{louboutin_devito_2019,komatitsch_unsplit_2007}. These implementations of standard numerical methods have enabled effective simulations of wave propagation equations in seismology \citep{moseley_deep_2020,roncoroni_synthetic_2021,mosser_stochastic_2020}.

Machine learning has recently shown significant potential for approximating the behavior of physical systems. For example, support vector machines (SVMs) have been used to solve ordinary and partial differential equations. Although this method was originally developed for classification tasks, extensions based on a least-squares formulation of the objective function have been proposed for the solution of differential equations \citep{mehrkanoon_approximate_2012,mehrkanoon_learning_2015}. 

\begin{figure*}
\centering 
\includegraphics[width=6.2 in]{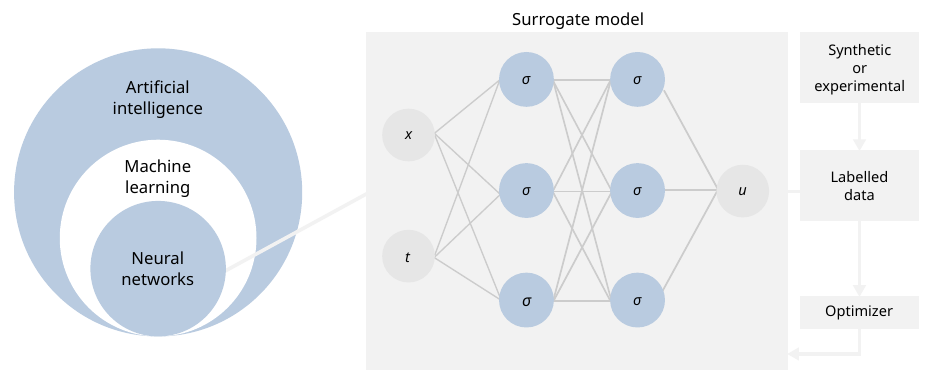}
\caption{Artificial intelligence hierarchy and surrogate model architecture. Neural networks represented as a subset of machine learning within the broader field of artificial intelligence (left). The right panel illustrates a fully-connected neural network used as a surrogate model, where inputs ($x$, $t$) propagate through two hidden layers of neurons with activation functions ($\sigma$) to produce output $u$. Model parameters are optimized iteratively using labeled synthetic or experimental data to minimize a loss function.}
\label{fig:surrogate}
\end{figure*}

Methods based on neural networks constitute a subset of machine learning, whose models are obtained from an artificial neural network with one or multiple processing layers (see \autoref{fig:surrogate}). These methods have shown potential in fields such as computer vision, natural language processing, and genomics \citep{lecun_deep_2015,goodfellow_deep_2016}. The fundamental architecture of a neural network consists of an input layer, an output layer, and an arbitrary number of hidden layers. In particular, in a fully connected neural network, neurons in adjacent layers are interconnected, while neurons within the same layer do not share connections. Furthermore, methods based on neural networks have emerged as an attractive tool to complement and extend conventional numerical solvers for partial differential equations.

These methods have often been implemented as data-driven approaches that do not explicitly account for the underlying physical system. In this sense, they are commonly regarded as black-box models, since the relationship between inputs and outputs is learned directly from data with limited use of prior physical knowledge. At the other end of the spectrum, white-box approaches rely on explicit physical modeling based on governing equations and established theoretical principles. Ignoring physical information is not always desirable, since physical knowledge can improve generalization and reduce data requirements, particularly when only limited observations are available. Physics-informed approaches aim to bridge these two perspectives by combining data-driven learning with physical modeling.

\begin{figure*} 
\centering
\includegraphics[width=6.3 in]{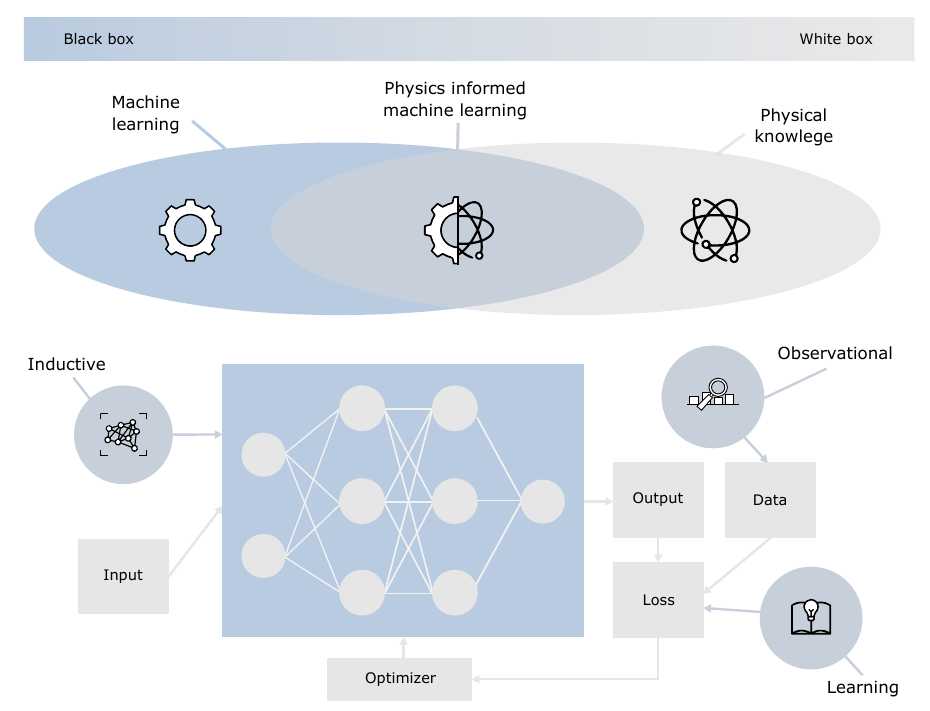}
\caption{Physics-informed machine learning taxonomy and knowledge 
incorporation mechanisms. Schematic representation of the PIML 
spectrum along a black box to white box continuum, in which purely 
data-driven machine learning (left) and physical-knowledge-based models 
(right) mark the two extremes, while PIML (center) integrates both. Within the neural-network training procedure, including 
input, network architecture, output, data, optimizer, and loss. Physical 
knowledge is incorporated through three complementary bias mechanisms: 
inductive bias (network architectures), 
observational bias (data), 
and learning bias (physics based loss).}
\label{fig:piml}
\end{figure*}

In the reviewed literature, the incorporation of physical information was considered into three categories: observational bias, inductive bias, and learning bias \citep{karniadakis_physics-informed_2021}. \autoref{fig:piml} illustrates this classification framework, depicting the  PIML spectrum as a continuum from purely data-driven to fully physics-based models, with the three bias mechanisms positioned according to where physical knowledge enters the training procedure.

\subsection*{Observational bias}
 
In observational approaches, neural networks are trained using either experimental observations of physical phenomena or synthetic data generated by standard numerical methods. In this setting, models can learn, for example, the solution of systems of partial differential equations directly from available data, without explicitly enforcing the governing equations during training \citep{li_neural_2020,li_fourier_2021}.

A motivation for this approach is the  construction of surrogate models that approximate solutions to the governing equations at a reduced computational cost while maintaining an acceptable level of accuracy. For example, \cite{moseley_deep_2020} trained two neural networks that were 20--500 times faster at inference than standard solvers. Similarly, \cite{das_surrogate_2018} proposed a surrogate modeling framework for 3D elastic wave propagation, using machine-learning regressors, such as Gaussian process, trained on synthetic seismograms generated from heterogeneous velocity models. Their surrogate models provided rapid and accurate seismogram predictions, enabling efficient likelihood evaluations and microseismic source localization without repeatedly solving the computationally expensive elastic wave equation.


A related family of methods is neural operator learning, in which the objective is to learn mappings between function spaces rather than between discrete input--output pairs. This allows approximation of the PDE solution operator across varying input conditions. However, these approaches typically require large amounts of training data, often generated through extensive numerical simulations, which can be computationally demanding. For example, \cite{yang_seismic_2021} showed that a Fourier Neural Operator trained on synthetic simulations generalized to new velocity models and source locations and could be evaluated on higher-resolution grids than those used during training. Also, \cite{yang_rapid_2023} applied a U-shaped Neural Operator (U-NO) to the 2D elastic wave equation, training on 20,000 synthetic finite-difference simulations, and demonstrated that once trained, the model could simulate full wavefields approximately 60 times faster than conventional finite-difference methods, while generalizing to velocity models, source locations, and mesh discretizations not seen during training. Extending this approach to three-dimensional settings, \cite{lehmann_multiple-input_2024} introduced the Multiple-Input Fourier Neural Operator (MIFNO), trained on 30,000 earthquake simulations with varying source characteristics and material properties. The model accurately reproduced wave arrival times and amplitude variations while achieving substantial computational speedups relative to standard numerical solvers. These results support the use of neural operators as surrogate models for wave propagation.

\subsection*{Inductive bias}

In inductive approaches, physical knowledge is incorporated directly into the structure of the neural network. Instead of learning from data alone, the model is designed so that its architecture reflects known properties of the physical system, such as the governing equations or spatial symmetries. Unlike approaches where physics enters as a penalty term in the loss function, here the network structure itself enforces the physics, meaning the model cannot produce solutions that violate the governing equations.

One example is the reformulation of wave propagation as a recurrent neural network (RNN), where each recurrent step corresponds to one time iteration of the discretized wave equation. In this formulation, the recurrent operators are determined directly by the governing physics, while the subsurface model parameters are updated through inversion. Because the network architecture implements the finite-difference method of the wave equation, the resulting model is constrained to represent physically consistent wave propagation. Building on this framework, \cite{wang_memory_2023} proposed a memory-efficient automatic differentiation strategy that reconstructs intermediate wavefields during backpropagation, allowing gradients to be computed through long simulations without storing the full computational history \citep{wang_memory_2023}.

Following this type of inductive approach, RNN formulations provide a framework for solving seismic inverse problems while explicitly incorporating the underlying wave physics. In these applications, subsurface properties become trainable parameters of the network, whereas the forward wave simulation is embedded directly in the architecture. For example, Lu and Zhang (2023) represented the finite-difference acoustic wave equation as a recurrent neural network and treated the velocity model as the parameter to be estimated. The inversion is performed by minimizing the mismatch between observed and simulated seismic data through backpropagation, allowing gradients to update the velocity model directly. To further exploit the causal nature of wave propagation, the authors proposed a time-by-time inversion strategy in which the velocity model updated at one time step serves as the initial model for the next. This approach reduces nonuniqueness, improves inversion stability, and decreases sensitivity to the initial velocity model \citep{lu_seismic_2023}. A related strategy was proposed by \citet{zhang_seismic_2024}, who integrated a data-driven velocity inversion network with an RNN-based forward modeling network that solves the acoustic wave equation. The predicted velocity model is used within the forward simulator, and the mismatch between observed and simulated seismic data is backpropagated to provide physics-based constraints during training. These studies demonstrate that integrating physical laws into neural network architectures can improve inversion stability, reduce nonuniqueness, and enhance the physical consistency of the recovered subsurface models.

Following this approach, \citet{ji_efficient_2024} developed a unified recurrent convolutional neural network framework applicable to electromagnetic, acoustic, and elastic wave equations. By implementing finite-difference as neural network architectures compatible with automatic differentiation and GPU acceleration, the proposed framework achieved computational speedups of up to two orders of magnitude while preserving the accuracy of the underlying numerical schemes.  

CNNs apply convolutional filters to the input data and are commonly implemented using two-dimensional convolutional layers, which makes them well suited for processing gridded data and learning spatial correlations. In elastic multiparameter inversion, CNN-based architectures have been used to reduce parameter cross-talk in reconstructed models \citep{dhara_elastic-adjointnet_2022}. Related CNN-based approaches have also been proposed for time-lapse monitoring. For example, \cite{gao_underground_2024} developed HydrogenNet, a CNN-based framework for detecting and characterizing underground hydrogen storage leakage from sparse seismic observations. The model directly estimates leakage location, mass, and volume from time-lapse seismic data \citep{gao_underground_2024}.

Inductive bias can also be encoded through the joint choice of data representation and model architecture. \cite{feng_intriguing_2022} showed that, after applying suitable integral transforms, transformed seismic measurements and subsurface properties exhibit an approximately linear relationship. Based on this observation, they proposed a shallow encoder-decoder architecture whose design assumes that the inversion mapping can be represented largely through near-linear interactions in the transformed domain. This assumption acts as an inductive bias, guiding the model toward a restricted class of solutions before any training data are observed. 
 
These forms of bias are not mutually exclusive and may be incorporated simultaneously within a single framework. For example, \cite{brandolin_pinnslope_2024} employed separate neural subnetworks to represent the seismic wavefield and the associated local slope field, thereby embedding prior knowledge about their distinct physical roles into the model architecture. At the same time, the local plane-wave differential equation was enforced through the training objective, introducing an additional physics-based inductive bias. This combination of architectural and physics-informed constraints yielded higher reconstruction quality and faster convergence than both plane-wave regularized least-squares interpolation and earlier single-network PINN formulations \citep{brandolin_pinnslope_2024}. The role of physics-based loss functions is discussed further in the following section.

A related form of architectural inductive bias was introduced by \cite{guo_parametric_2024}, who represented P-wave and S-wave velocity models through separate convolutional neural networks within an elastic full waveform inversion framework. Rather than updating the velocity fields directly, the inversion optimizes the network parameters that generate them. This parameterization imposes spatial structure on the admissible models through the convolutional architecture, thereby acting as an implicit regularizer. The resulting representation was reported to reduce parameter cross-talk and improve robustness to local minima relative to conventional elastic FWI \citep{guo_parametric_2024}.

By embedding assumptions about the governing physics, spatial structure, or parameter interactions directly into the network architecture, these approaches reduce the solution space and improve the stability of the learning process.

\subsection*{Learning bias}

\begin{figure*}
\centering
\includegraphics[width=6.4in]{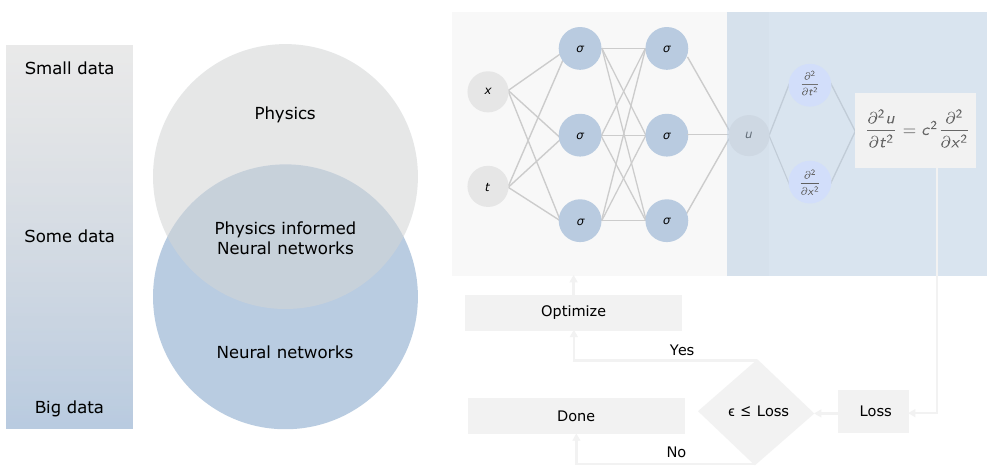}
\caption{Physics-informed neural networks: data requirements and architecture. The relationship between physical knowledge and data: purely data-driven neural networks rely on large datasets, whereas PINNs incorporate physical information, making them applicable with limited data. The right panel illustrates the PINN architecture, where the network predicts the field $u(x,t)$ from inputs ($x$, $t$), and spatial and temporal derivatives are embedded into the loss function via the wave equation $\frac{\partial^2 u}{\partial t^2} = c^2 \frac{\partial^2 u}{\partial x^2}$, ensuring physical consistency throughout the optimization process.}
\label{fig:pinns_scheme}
\end{figure*}

In learning-based approaches, physical knowledge is incorporated directly into the training process through the loss function. Among these methods, PINNs represent a widely used framework for solving partial differential equations (see \autoref{fig:pinns_scheme}). Their relevance lies in the combination of observational information with governing physical laws, which is especially valuable in seismological problems where data are often sparse, noisy, or incomplete.

PINNs approximate the solution of a differential equation by training a neural network that minimizes a loss function composed of the PDE residual together with terms enforcing initial and boundary conditions. Formally, consider a general PDE defined over a domain $\Omega$ with boundary $\Gamma$:
\[
    \mathcal{N}[u(\mathbf{x}, t); \lambda] = 0, \quad \mathbf{x} \in \Omega, \quad t \in [0, T],
\]
where $\mathcal{N}$ is a differential operator parameterized by $\lambda$, and $u(\mathbf{x}, t)$ is the unknown solution. The boundary and initial conditions are given by:
\[
    \mathcal{B}[u(\mathbf{x}, t)] = 0, \quad \mathbf{x} \in \Gamma, \quad t \in [0, T],
\]
and
\[
    u(\mathbf{x}, 0) = u_0(\mathbf{x}), \quad \mathbf{x} \in \Omega.
\]

The total loss function is decomposed into three terms:
\[
    \mathcal{L} = \mathcal{L}_{\Omega} + \mathcal{L}_{\Gamma} + \mathcal{L}_{\text{IC}},
\]
where $\mathcal{L}_{\Omega}$ enforces the governing PDE within the domain, $\mathcal{L}_{\Gamma}$ enforces the boundary conditions on $\Gamma$, and $\mathcal{L}_{\text{IC}}$ enforces the initial conditions at $t = 0$. This formulation encourages the model to produce solutions consistent with the underlying physics while reducing dependence on large training datasets, as summarized in Algorithm~\ref{alg:pinn}.

\begin{algorithm}[ht]  
    \caption{\small\sffamily Physics-Informed Neural Network for Solving PDEs}  
    \label{alg:pinn}  
    \begingroup
    \small\sffamily
    \textbf{Inputs:}\\ 
    Set of sampling points in the domain \( \Omega \) with the governing PDE\\
    Initial conditions\\
    Boundary conditions \\  
    \textbf{Outputs:} \\ 
    Approximation of the solution \( u(\mathbf{x}, t) \) \\
    In the case of inverse problems, estimated parameters \( \lambda \)  
    \begin{algorithmic}  
    \State \textbf{Initialize parameters:} \( \theta \) (and \( \lambda \) if inferred)
    \State \textbf{Define hyperparameters:} \( \text{tol} \), \( \text{maxiter} > 0 \), \( \alpha_k \) 

    \For{optimization method in \{Adam, L-BFGS\}}  
        \State \textbf{Initialize iteration counter:} \( k = 0 \)  
        \While{$k < \text{maxiter}$ \textbf{and} $L(\theta^{(k)}) > \text{tol}$}  
            \State \textbf{Evaluate:} \( u(\mathbf{x}, t; \theta^{(k)}) \)  
            \State \textbf{Compute total loss:}  
            \[
                L(\theta^{(k)}) = L_{\Omega} + L_{\Gamma} + L_{\text{IC}}
            \]  
            \State \textbf{Update:}  
            \[
                \theta^{(k+1)} = \theta^{(k)} - \alpha_k \nabla L(\theta^{(k)})
            \]  
            \State \textbf{Increment iteration:} \( k = k + 1 \)  
        \EndWhile  
    \EndFor  
    \State \textbf{Return} Trained model: \( u(\mathbf{x}, t; \theta) \)  
    \end{algorithmic}  
    \endgroup
\end{algorithm}

Early foundations for this approach were introduced by \citet{dissanayake_neural-network-based_1994} (\citeyear{dissanayake_neural-network-based_1994}), who reformulated the solution of differential equations as an unconstrained optimization problem using neural networks as function approximators. \cite{lagaris_artificial_1998} subsequently refined this formulation by proposing trial solutions composed of two parts: one designed to satisfy initial or boundary conditions analytically, and another represented by a neural network. Building on these foundations, \cite{raissi_physics-informed_2019} reintroduced and extended the methodology using modern deep learning techniques, leading to the development of PINNs as currently understood.

Two approaches are considered. The first focuses on data-driven solutions for direct problems, aiming to infer the hidden state $u(\mathbf{x}, t)$ from known model parameters $\lambda$. The second focuses on parameter identification in inverse problems, seeking the values of $\lambda$ that best fit the observed data. For each case, two models are implemented: the continuous-time model, which approximates a space-time function, and the discrete-time model, which uses Runge-Kutta methods for time integration.

\begin{figure*}[ht!]
\centering
\includegraphics[width=6.2in]{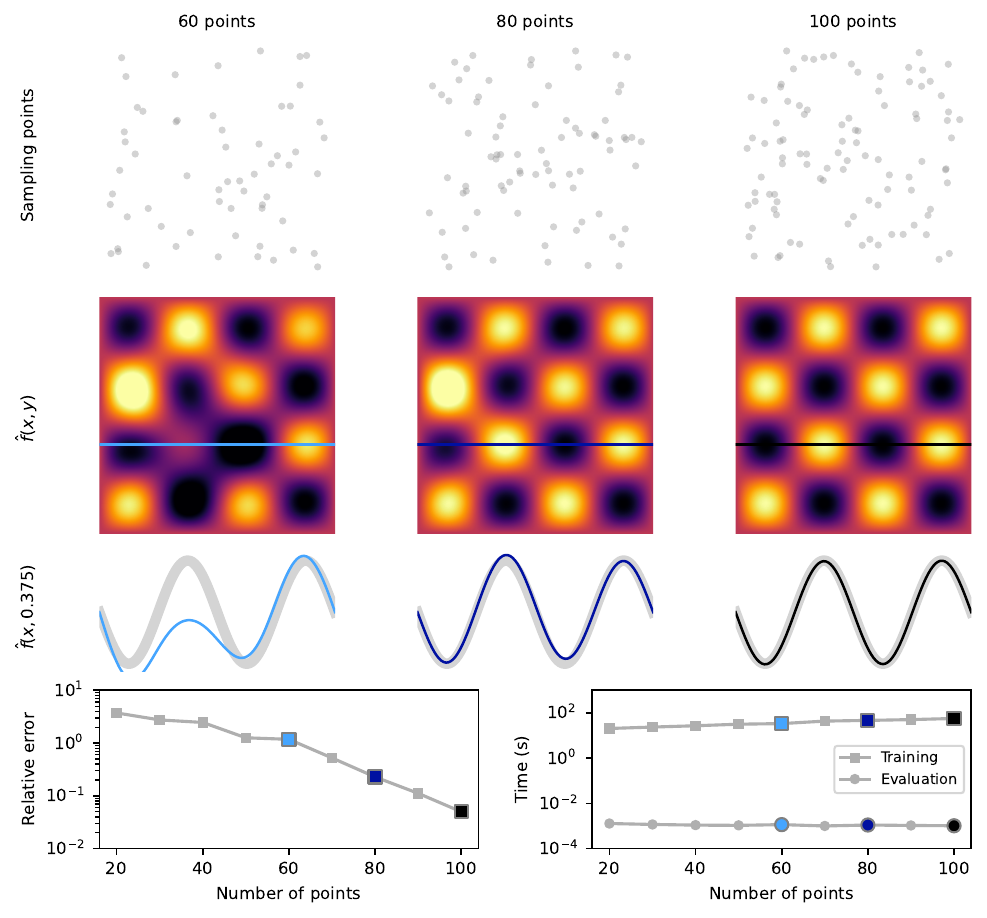}
\caption{PINN performance for the 2D Helmholtz problem with different training point densities. (Top row) Random sampling distributions used during training, corresponding to 60, 80, and 100 points. (Second row) Predicted fields $\hat{f}(x,y)$ for each case. (Third row) Horizontal cuts at $y = 0.375$, comparing the PINN prediction (blue, dark blue, and black lines) with the analytical solution (gray line). (Bottom row) Convergence plots showing the relative error (left) and computational times (right), both on a logarithmic scale.}
\label{fig:pinn_helmholtz2D_convergence}
\end{figure*}

The PINNs performance in solving the same two-dimensional Helmholtz problem described in \autoref{eq:helmholtz2D} is shown in \autoref{fig:pinn_helmholtz2D_convergence}. The optimized hyperparameters in the work of \cite{escapil-inchauspe_hyper-parameter_2023} and \textit{hard-constrains} were used for Dirichlet boundary conditions.
 
In this case, the neural network was defined as:
\[
\hat{f}_\theta(x, y) = x(1 - x)\, y(1 - y)\, f_\theta(x, y),
\]
where \(f_\theta(x, y)\) represents the unconstrained output of the neural network, that is, the function that the PINN learns during training.

The factor \(x(1 - x)y(1 - y)\) guarantees that \(\hat{f}_\theta(x, y)\) vanishes on the boundary of the domain $[0,1]^2$, exactly satisfying the homogeneous boundary conditions specified in the \autoref{eq:helmholtz2D}.

It is observed that, by increasing the number of training points, the network improves its ability to approximate the analytical solution, reducing the relative error. This behavior results from better spatial coverage of the domain, allowing for a more detailed capture of the solution's variations. In contrast, the model's evaluation time, once trained, remains practically constant, regardless of the number of points used. This is because the inference phase depends solely on the size of the evaluation point set and the network architecture, and not on the prior optimization process. This result demonstrates the efficiency of the PINN model in making predictions after training is complete. However, the training time reaches values on the order of \(10^{1}\,\text{s}\), which is two orders of magnitude greater than the maximum time recorded with the finite difference method in the previous example.

By embedding physical laws directly into the learning process, PINNs provide a unified framework for obtaining physically consistent solutions that integrates data and PDE-based constraints within a single optimization problem. In seismological applications, this strategy has shown potential for improving the reconstruction of aliased seismic data and large gaps compared with purely data-driven networks. An example of this learning bias strategy applied to inverse problems is the work of \cite{xu_physics_2019}, who encoded the acoustic wave
equation directly into the loss to perform velocity inversion from synthetic seismic data. Their approach yielded accurate results compared with FWI on two-dimensional problems, demonstrating that the physics-informed loss can effectively constrain the solution space in parameter estimation tasks  \citep{xu_physics_2019}.

Nevertheless, important challenges remain, including the representation of highly oscillatory seismic signals and wavefields, which are often difficult for conventional PINNs due to their spectral bias toward low-frequency components. Architectural modifications such as positional encoding have been shown to improve the learning of multi-scale seismic features and accelerate convergence in physics-informed frameworks \citep{brandolin_pinnslope_2024}. Another practical limitation of conventional PINNs is the need to retrain the network for each new velocity model. To address this issue, \cite{zou_seismic_2024} proposed a velocity-encoded PINN (VE-PINN) framework in which geological feature parameters, such as layer velocities, interface depths, and source locations, are incorporated directly into the network inputs. This enables the model to generalize across multiple layered velocity configurations and source positions within the training range without requiring retraining for each new scenario \citep{zou_seismic_2024}.

Several extensions of the original PINN framework have been proposed. \cite{kharazmi_variational_2019} introduced variational PINNs, in which the physics loss is formulated using the weak form of the governing equations: the neural network is combined with analytical test functions, and quadrature points are used to estimate the resulting integrals. Their formulation yielded similar or improved accuracy relative to standard PINNs while requiring fewer collocation points. However, most such extensions have not yet been widely applied to wave propagation problems. The growing number of such variants underscores the need for reliable, reusable baseline implementations. Without reproducible foundations, each new extension must be evaluated.

\subsection*{Replication of the PINN Framework}

Reproducibility is a fundamental practice in science, as it allows results to be verified and builds confidence in published findings \citep{perkel_challenge_2020}. Scientific replication refers to the independent re-implementation of a published method from its original description, with the goal of confirming that reported results can be obtained under equivalent conditions \citep{drummond_replicability_2009}. Unlike reusing the original code and data, replication requires rebuilding the computational pipeline from scratch, which provides stronger evidence of a method's correctness and generalizability.

This is particularly important in computational science, where results can be sensitive to software dependencies, numerical precision, and implementation decisions. However, researchers are given few incentives to invest time in replication practices such as documenting code, creating test suites, and archiving datasets. This is a growing concern, given how rapidly scientific tools and software environments evolve \citep{barba_defining_2022}.

Despite the growing number of PINN extensions, reproducibility of the original framework remains a practical concern: the implementation by \cite{raissi_physics-informed_2019} was developed in TensorFlow without unspecified library versions and is incompatible with
current Python environments. The emergence of actively maintained frameworks, including NeuroDiffEq, DeepXDE, and PINNs-Torch, makes re-implementation in modern tooling both feasible and valuable as a foundation for future work \citep{chen_neurodiffeq_2020,lu_deepxde_2021,bafghi_pinns-torch_2023}.

To this end, we re-implemented the original PINN formulations using PyTorch, preserving the original hyperparameters (learning rates, network depth and width, and number of sampling points) and architectural choices: feed-forward networks with hyperbolic tangent activations and Xavier initialization. Collocation points were drawn via Latin Hypercube Sampling to improve input domain coverage, and PDE residuals were evaluated through automatic differentiation. Training followed a two-stage procedure combining the Adam optimizer for initial convergence and L-BFGS for fine-tuning. All experiments were run on an NVIDIA GeForce RTX A2000 GPU, and accuracy was assessed using the relative $\text{L}_2$-norm against analytical solutions.

The resulting scripts are available at \url{https://github.com/oscar-rincon/Replication-PINNs}. The organization and documentation of the repository were guided by the reproducibility recommendations and checklist proposed in ``Challenge to scientists: does your ten-year-old code still run?'' \citep{perkel_challenge_2020}. To support long-term computational reproducibility, the repository includes a dedicated Conda environment, dependency specifications, and a master execution script that automates the complete workflow.  

The replication results are summarized in \autoref{tab:comparison_errors}.  For the forward problems, the experiments considered the continuous Schrödinger equation and the discrete Allen--Cahn equation. In both cases, the performance metric was the relative $L_2$ error between the predicted and reference solutions. The PyTorch re-implementation reproduced the original results successfully and obtained lower relative $L_2$ errors than those reported in the original study. For the inverse problems, the experiments involved the continuous Navier--Stokes equations and the discrete Korteweg--de Vries equation. In these cases, the evaluation metric corresponded to the percentage error in the identified parameters $\lambda_1$ and $\lambda_2$, using both clean and noisy datasets. For the Navier--Stokes equations, the replicated implementation achieved lower parameter errors across all experiments, both for clean and noisy data. In the Korteweg--de Vries equation, the replicated results also improved the parameter estimation errors for the clean-data case. However, under noisy conditions, the reproduced errors increased compared to the original implementation, particularly for $\lambda_1$ and $\lambda_2$, indicating a higher sensitivity to noise in this discrete inverse setting.

\begin{table} 
\centering
\sffamily\small
\caption{Comparison of errors between the original study and the replicated 
results for PINNs.}
\begin{tabular}{lcc}
\hline
\textbf{Model / Parameter} & \textbf{Original} & \textbf{Replication} \\
\hline

\multicolumn{3}{l}{\textbf{Forward problems} } \\
\hline

\multicolumn{3}{l}{\textbf{Continuous forward - Schrödinger equation}} \\
Relative $L_2$ error & 0.00197 & 0.001307 \\

\multicolumn{3}{l}{\textbf{Discrete forward - Allen--Cahn equation}} \\
Relative $L_2$ error & 0.00699 & 0.001307 \\
\hline

\multicolumn{3}{l}{\textbf{Inverse problems  (\% error)}} \\
\hline

\multicolumn{3}{l}{\textbf{Continuous inverse - Navier--Stokes equations}} \\
$\lambda_1$ (clean data) & 0.078\% & 0.007\% \\
$\lambda_2$ (clean data) & 4.670\% & 1.864\% \\
$\lambda_1$ (noisy data) & 0.170\% & 0.029\% \\
$\lambda_2$ (noisy data) & 5.700\% & 3.290\% \\

\multicolumn{3}{l}{\textbf{Discrete inverse - Korteweg--de Vries equation}} \\
$\lambda_1$ (clean data) & 0.023\% & 0.004\% \\
$\lambda_2$ (clean data) & 0.006\% & 0.005\% \\
$\lambda_1$ (noisy data) & 0.057\% & 0.119\% \\
$\lambda_2$ (noisy data) & 0.017\% & 0.050\% \\
\hline

\end{tabular}
\label{tab:comparison_errors}
\end{table}

The growing availability of open-source frameworks is expected to facilitate broader adoption and more systematic benchmarking of PINNs in geophysical applications. A natural direction for future work is the replication of these experiments using alternative frameworks better suited for high-performance scientific computing. In the Python ecosystem, JAX \citep{jax2018github} offers a compelling alternative
to PyTorch, providing composable function transformations, just-in-time compilation via XLA, and native support for automatic differentiation and vectorization, all of which are advantageous for PINN training.

Beyond Python, Julia-based implementations, particularly NeuralPDE \citep{zubov_neuralpde_2021}, warrant evaluation for their native support for automatic differentiation, high-performance numerical routines, and elimination of the two-language problem commonly ncountered
in Python-based workflows \citep{bezanson_julia_2014}. Together, these frameworks represent promising directions for scaling PINNs toward realistic geophysical scenarios.
 
\section*{Hybrid approaches}

Hybrid formulations reflect ongoing efforts to improve the scalability and practical applicability of machine learning methods in seismological problems. Rather than replacing established numerical solvers, these approaches combine methods in order to exploit the complementary strengths of both. For example, numerical methods provide physically grounded and accurate descriptions of wave propagation, whereas machine learning methods can reduce computational cost, improve robustness to noise, and mitigate the effects of incomplete observations.

These hybrid strategies have been reported in the reviewed literature. For example, \cite{zhang_autoencoded_2023} integrated an autoencoder-based feature extraction framework into elastic wave-equation traveltime inversion. Instead of relying on manually picked first-arrival traveltimes, the method employs latent-space representations of seismic records within the inversion objective function. By combining data-driven feature extraction with physics-based wave-equation modeling, the approach improved the stability and robustness of near-surface velocity tomography in the presence of low signal-to-noise ratios and inconsistencies in first-arrival measurements \citep{zhang_autoencoded_2023}. Similarly, \cite{fang_shear-wave_2024} integrated poroelasticity theory with a deep neural network to predict shear-wave velocity from well-log measurements in tight reservoirs, demonstrating improved predictive accuracy relative to conventional rock-physics-based approaches \citep{fang_shear-wave_2024}.

Hybrid strategies have also been applied to forward wave modeling, where neural networks are trained on outputs of classical solvers and subsequently used as fast surrogates. \cite{roncoroni_synthetic_2021} trained a neural network on finite-difference simulations to generate acoustic synthetic seismograms from 1D velocity models, achieving a prediction speedup of approximately 27 times relative to the finite difference solver \citep{roncoroni_synthetic_2021}. A related approach was proposed for wave propagation in fluid-saturated rocks, where deep neural networks were trained as surrogates for the Biot--squirt (BISQ) model. One network learned the mapping between rock-physics parameters and wave-dispersion and attenuation attributes generated through plane-wave analysis of the BISQ equations, achieving relative errors below 3\%, while a second network learned to approximate second-order finite-difference solutions from first-order finite-difference wavefields, thereby accelerating high-accuracy simulations \citep{xiong_deep-neural-networks-based_2022}. Similarly, \cite{hateleyDeepLearningSeismic2020} trained convolutional neural networks on synthetic wavefields generated by the frozen Gaussian approximation, a computationally efficient semiclassical solver, and applied them to seismic interface and pocket detection, with the additional finding that networks generalized across solvers when traveltime information was preserved. Together, these examples illustrate how standard solvers can be used to generate training data, with the network subsequently acting as a surrogate at inference time. 


By executing classical finite-difference iterations within a deep learning framework, and exploiting GPU-accelerated libraries such as cuDNN, the discretized PDE solver itself becomes the network, with no training on labeled data required. This design has been reported to reduce computation time by roughly 24\% relative to conventional implementations while remaining applicable across different wave equation types \citep{ji_efficient_2024}. Such an approach illustrates how deep learning infrastructure can serve as an efficiency layer on top of established numerical methods, without modifying the underlying physics.

Future research could further expand this hybrid perspective by integrating machine learning methods with established numerical simulation frameworks such as SPECFEM or Devito. In this context, machine learning could be used as a preprocessing stage. For example, to generate improved initial models, denoise observations, or extract relevant features before numerical simulation. Also, to support surrogate modeling of selected workflow components, adaptive parameter estimation, and accelerated inversion updates.

\begin{figure}
\centering
    \includegraphics[width=2.9 in]{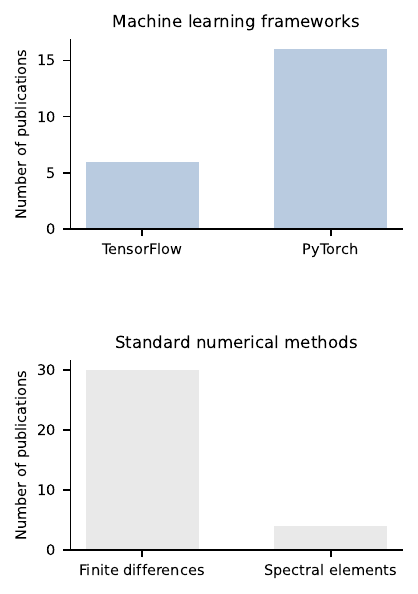}
\caption{Distribution of publications according to the machine learning frameworks and standard numerical methods used in the reviewed studies. PyTorch was more frequently employed than TensorFlow in machine learning implementations, while finite-difference methods were more commonly used than spectral-element methods among standard numerical approaches.}
\label{fig:frameworks_methods}
\end{figure}

Beyond computational advantages, machine learning methods provide a relatively accessible framework for researchers from related disciplines. Their implementation is supported by general-purpose libraries with extensive documentation and active user communities, while their higher-level formulation reduces the need to directly implement low-level numerical discretization methods commonly required in standard finite differences or spectral element methods. This accessibility, combined with the flexibility to incorporate data, positions machine learning as a bridge for multidisciplinary collaboration in seismological research, where expertise in signal processing, geology, and applied mathematics increasingly converges. The distribution of machine learning frameworks and standard numerical methods identified in the reviewed studies is summarized in \autoref{fig:frameworks_methods}.

\section*{Applicability in inverse problems}

Inverse problems are a particularly relevant setting for evaluating the potential advantages of machine learning methods, since they often require repeated forward simulations and can therefore become computationally demanding. In the reviewed literature, machine learning approaches are commonly reported as complementary tools to standard numerical techniques rather than direct replacements.

Some research efforts focused on integrating learned components into existing inversion frameworks. \cite{yang_revisit_2021} proposed FWIGAN, an unsupervised physics-informed inversion framework that combines the acoustic wave equation with a Wasserstein Generative Adversarial Network with Gradient Penalty (WGAN-GP). Rather than replacing the forward solver, the method uses a CNN-based critic to measure discrepancies between observed and simulated seismic data in a distributional sense, improving robustness to noise and poor initial models while reducing sensitivity to local minima \citep{yang_revisit_2021}. Similarly, \cite{mosser_stochastic_2020} combined a GAN-based geological prior with the adjoint solution of the acoustic wave equation, reducing the inversion to a lower-dimensional latent space and enabling efficient Bayesian posterior sampling of subsurface models \citep{mosser_stochastic_2020}. These approaches illustrate the potential of GAN approaches to complement traditional inversion methods while preserving physical consistency.

Studies are also moving toward replacing specific numerical components with learned operators. \cite{fu_seismic_2019} replaced the finite-difference forward operator in FWI with a shallow feedforward neural network, reducing the total inversion time from 6933.8 s to 4832.4 s and decreasing the number of conjugate-gradient iterations from 643 to 482 in Marmousi-model experiments. The learned surrogate provided a computationally efficient approximation of the forward operator while preserving the main features of the recovered velocity model \citep{fu_seismic_2019}. A different strategy was proposed by \cite{alfarhan_robust_2024}, who used a neural network to approximate the inverse Hessian in FWI, a generally computationally expensive procedure. The network acts as an approximate inverse Hessian, enhancing the FWI gradient at each iteration and providing a more effective update direction, which leads to faster convergence and improved subsurface models. The approach was validated on both synthetic and real data, making it one of the reviewed studies to demonstrate applicability beyond synthetic benchmarks \citep{alfarhan_robust_2024}.

A parallel line of work pursued fully observational formulations in which the entire inversion is replaced by a trained network. \cite{wang_velocity_2018} trained a fully convolutional network on synthetic seismic traces to predict P-wave velocity models directly, demonstrating rapid velocity reconstruction and arguing that the direct model-domain prediction framework reduced sensitivity to cycle skipping issues commonly encountered in conventional FWI while requiring negligible computational cost at inference time. \cite{ren_physics-based_2020} extended this integration further by embedding wave-equation-based forward modeling directly into neural network cells within SWINet, a PyTorch-based implementation in which automatic differentiation and the Adam optimizer accelerate FWI convergence while reducing computational cost relative to standard implementations. In a related direction, \cite{yang_seismic_2021} trained a Fourier Neural Operator (FNO) as a surrogate solver for the acoustic wave equation using an ensemble of numerical simulations generated from diverse velocity models and source locations. The resulting model generalized to previously unseen velocity structures and source configurations without retraining and enabled gradient-based FWI through automatic differentiation. In tomography experiments, the FNO-based workflow achieved inversion results comparable to those obtained with spectral-element simulations and adjoint-state methods while reducing the cost of a single iteration from approximately 100 s to 1 s. These methods show that machine learning can substantially reduce the cost of seismic inversion.

PINNs offer a complementary perspective by incorporating governing equations directly into the training process, providing a framework for parameter estimation that maintains physical consistency without requiring large labeled datasets. \cite{karimpouli_physics_2020} demonstrated this by applying both Gaussian process regression and PINNs to the 1D seismic wave equation, showing that both methods can invert unknown wave parameters such as P- and S-wave velocities and density, with PINNs achieving higher inversion accuracy and GP providing uncertainty estimates through its probabilistic formulation. More recent PINN formulations have moved beyond single network architectures: \cite{zhang_seismic_2023} employed separate networks to estimate the displacement field and to infer the velocity and density fields, improving the representational capacity of the approach.

Beyond accuracy, robustness to measurement noise and computational efficiency at inference time are critical criteria for practical deployment. \cite{gelboim_encoder-decoder_2023} demonstrated that a 3D convolutional encoder--decoder network can reconstruct subsurface velocity models from seismic data contaminated by both white Gaussian and field noise, achieving a structural similarity index of 0.9143 at a signal-to-noise ratio of 10,dB. In a related setting, \cite{lahivaara_deep_2022} trained a convolutional neural network on synthetic seismic data generated from coupled elastic--poroviscoelastic wave simulations to estimate groundwater-related properties, including water table levels and aquifer storage volumes, from noisy seismic recordings. Their results showed accurate predictions even in the presence of substantial measurement noise, highlighting the robustness of convolutional architectures across different geophysical inverse problems. Complementing these results, \cite{cao_near-real-time_2020} combined Helmholtz wave-equation inversion with a pretrained mixture density network to transform ambient seismic noise recordings into 3D near-surface velocity models with uncertainty estimates. The proposed workflow produced velocity models within seconds and demonstrated the potential for near-real-time subsurface characterization using only short-duration ambient noise recordings.

The reviewed studies show a clear trend toward hybrid formulations that reduce the computational burden of seismic inversion while preserving physical consistency. Machine learning is increasingly embedded in the core of inversion workflows as both a surrogate for costly forward simulations and a tool for direct parameter estimation, reducing inference time from hours to seconds while preserving the physical structure of the problem.

\section*{Toward scalable machine learning models}

The use of experimental data introduces additional challenges, including measurement noise, increased problem dimensionality, and incomplete knowledge of material properties. Consequently, an important research gap concerns the integration of machine learning models into uncontrolled seismological settings, understood as scenarios that extend beyond the simplified benchmark problems commonly adopted for methodological validation. Addressing this gap is essential for translating current advances into practical and reliable seismological applications.

\begin{figure*}
\centering
    \includegraphics[width=6.2 in]{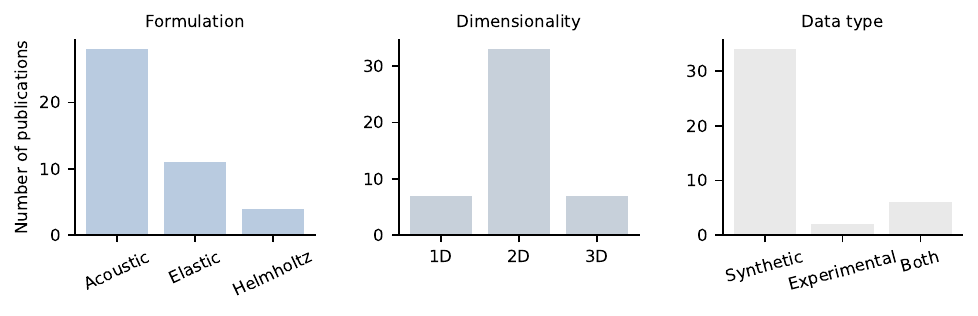}
\caption{Methodological trends in the reviewed literature. The bar plots show the distribution of the reviewed studies according to the governing wave equation, problem dimensionality, and data type (synthetic or experimental), highlighting prevailing modeling choices and current limitations in the literature.}
\label{fig:equation_dimensionality_data_type}
\end{figure*}

One of the main observations from this review is the limited extent to which the proposed methods have been validated using experimental and/or field data. Several recurring characteristics are identified across the reviewed studies, including the widespread use of simplified mathematical formulations, predominantly based on acoustic wave equations, and reduced problem dimensionality. In addition, most investigations rely on synthetic datasets generated from idealized models. Although this facilitates controlled methodological evaluation, it restricts the assessment of model robustness under realistic conditions. As illustrated in \autoref{fig:equation_dimensionality_data_type}, the majority of the reviewed works focus on two-dimensional formulations and synthetic datasets, whereas only a limited number report validation using experimental or field observations. Among these, \cite{park_low-frequency_2024} validated a modified U-Net for low-frequency reconstruction in marine seismic data using field data from the North Viking Graben. Although the study addressed seismic data enhancement rather than velocity inversion directly, it demonstrated the applicability of deep learning beyond synthetic benchmarks while remaining within a two-dimensional acoustic framework \citep{park_low-frequency_2024}.

Overall, the reviewed literature shows that neural network-based methods are increasingly applied to wave propagation problems in seismology. However, their use remains concentrated in simplified settings characterized by acoustic formulations, reduced dimensionality, and synthetic datasets. Comparatively fewer studies address elastic wave propagation, large-scale domains, or validation using experimental data. These observations highlight important directions for future work, particularly the extension of learning-based methods toward more realistic physical models and validation scenarios.

The numerical modeling of high-dimensional problems governed by partial differential equations faces a fundamental challenge known as the curse of dimensionality \citep{bellman_dynamic_1957}, where memory requirements and computational costs increase exponentially with the number of dimensions. In seismology, three-dimensional (3D) wave propagation simulations are often required to accurately represent complex geological structures. Although standard numerical methods provide high accuracy, their application to high-dimensional systems remains computationally expensive, especially when fine-scale features need to be resolved. Therefore, reducing the computational cost of 3D simulations remains a major challenge.

Although machine learning approaches have demonstrated potential for reducing this cost, most current applications remain limited to one- or two-dimensional domains, as noted by \cite{blechschmidt_three_2021}. Similarly, in physics-based deep learning inversion, scalability to large-scale two-dimensional and especially three-dimensional problems is constrained by GPU memory requirements \citep{ren_physics-based_2020}. To address these limitations, \cite{lehmann_fourier_2024} proposed a Fourier Neural Operator framework for three-dimensional seismic wave propagation. A more recent and direct approach to this challenge was reported by \cite{zou_deep_2023}, who trained a U-shaped neural operator to solve the 3D elastic wave equation in the frequency domain. By working in the frequency domain rather than the time domain, the memory requirements were reduced by a factor of 40 relative to an equivalent time-domain formulation. The trained operator achieved forward simulation speeds roughly 100 times faster than a spectral element method, and when used for FWI via automatic differentiation, the computation time was reduced by a factor of 350. However, the training stage required over 30,000 numerical simulations, illustrating that significant computational investment and domain size restrictions remain even in the most advanced current approaches \citep{zou_deep_2023}. 

Similar challenges are observed in inversion methods. For example, \cite{gelboim_encoder-decoder_2023} proposed a 3D convolutional encoder--decoder network for seismic velocity inversion containing approximately 99 million trainable parameters. The authors introduced a dedicated dimensionality-reduction strategy to make the problem computationally feasible \citep{gelboim_encoder-decoder_2023}. While machine learning can substantially accelerate seismic modeling and inversion, extending such approaches to realistic three-dimensional domains continues to require considerable computational resources and specialized strategies for managing memory and training costs.

Another important limitation concerns the frequency content of the learned solutions. Neural networks, and PINNs in particular, exhibit spectral bias, meaning that low-frequency components are learned more easily than high-frequency features \citep{song_simulating_2023}. This limitation is especially relevant in seismic wave modeling due to the multifrequency nature of seismic wavefields. To mitigate this issue, \cite{song_simulating_2023} incorporated Fourier features into the PINN architecture, enabling the simulation of multifrequency wavefields governed by the Helmholtz equation with improved accuracy and efficiency. In this context, Fourier feature PINNs have demonstrated a reduction in wavefield prediction error over plain PINNs, with the $L_{2}$ norm of the scattered wavefield difference dropping from 687.37 to 13.20 when using the proposed Fourier feature embedding with a theoretically calibrated wavenumber sampling range \citep{song_simulating_2023}. Likewise, \cite{song_high-frequency_2022} demonstrated that Fourier Neural Operators can learn a mapping between low-frequency wavefields (5--12 Hz) and higher-frequency wavefields (13--30 Hz),  predicting high-frequency components from low-frequency inputs while maintaining dispersion-free solutions and achieving nearly two orders of magnitude faster wavefield generation than finite-difference modeling \citep{song_high-frequency_2022}.

Challenges are also associated with the required input data. For example, \cite{xu_physics_2019} reported that their PINN-based velocity inversion approach, although competitive with FWI on synthetic two-dimensional datasets, remains limited to wavefield snapshots as input and has not yet been extended to three-dimensional domains. This example illustrates how data requirements and scalability constraints continue to limit the applicability of current methodologies \citep{xu_physics_2019}. A complementary perspective on data requirements is provided by \cite{campos_empirical_2019}, who systematically studied how acquisition parameters, specifically the number of seismic sources and the peak frequency of the source signal, affect the quality of velocity model estimation by a fully convolutional network. Their results showed that performance does not improve monotonically with either more sources or higher frequencies, and that different velocity models may require different acquisition configurations to achieve the best results. The study was restricted to relatively small two-dimensional synthetic models, leaving the applicability of these findings to larger and more geologically realistic settings uncertain \citep{campos_empirical_2019}.

A further practical challenge is the quantification of prediction uncertainty. Most machine-learning inversion methods produce a single estimate without a measure of confidence, which limits their applicability in practice. To address this issue, \cite{zhang_physics-guided_2023} incorporated Bayesian layers into a physics-guided CNN for seismic impedance inversion, enabling probabilistic predictions and the generation of uncertainty maps alongside the inverted model. This framework provides statistically grounded estimates of inversion uncertainty and improves the interpretability of the recovered subsurface properties \citep{zhang_physics-guided_2023}.

Wave modeling in seismology is progressively moving toward the integration of conventional numerical methods with machine learning approaches. While traditional numerical solvers remain essential for accurate physical simulation, machine learning methods provide new opportunities to reduce computational costs through tools such as neural networks as function approximators, automatic differentiation, and GPU-based optimization. An example is the work of \citet{huang_applying_2022}, who implemented a finite difference solver for the wave equation within the PyTorch framework, allowing automatic differentiation to be used for gradient computation in FWI. This approach shows that deep learning platforms can serve as a basis for seismic simulation without replacing the underlying physics. The combination of both approaches has the potential to improve seismic simulation and inversion workflows, including applications related to time-lapse monitoring and parameter estimation from sparse seismic observations \citep{gao_underground_2024}.

Taken together, the reviewed studies indicate that simplified formulations, restricted dimensionality, dependence on synthetic datasets, GPU memory limitations, spectral bias, and limited uncertainty quantification remain the principal obstacles to the deployment of machine learning methods in operational seismological workflows.

\section*{Strengths and limitations of the present review}

This review offers a structured overview of physics-informed machine learning methods applied to seismic wave propagation, covering both forward modeling and inverse problems. The scoping review format is appropriate for a field that is expanding rapidly, as it allows the mapping of current approaches and the identification of research gaps without requiring a formal quality assessment of individual studies.

A practical strength of this work is its attention to reproducibility. The database queries, filtering steps, and data charting code are made publicly available, allowing readers to trace the selection process and verify the descriptive results. In addition, the re-implementation of the original PINN framework in PyTorch provides a working and documented baseline that can serve as a starting point for future studies.

Some limitations should be considered when interpreting the findings. The search was restricted to OpenAlex and Scopus, which, although broadly representative, may not capture all relevant publications. Since both databases are continuously updated, repeating the same queries at a later date may return a different set of records.

A further limitation concerns the comparisons reported in the included studies. As noted by \cite{mcgreivy_weak_2024}, performance comparisons between machine learning and numerical methods depend strongly on the choice of baseline, and these choices are not always appropriate or clearly justified. The inclusion criteria required that studies report quantitative comparisons or qualitative assessments of computational efficiency, but other relevant aspects were not evaluated systematically, such as whether methods were compared at equivalent accuracy levels or against suitable state-of-the-art numerical solvers \cite{cardenoBenchmarkingPhysicsinformedNeural2026}. As a result, the studies differ considerably in how comparisons were conducted, which limits the conclusions that can be drawn about relative performance across the literature.

No formal assessment of methodological quality was carried out, which is consistent with the scoping review methodology but means that the findings should be read as a map of current practices rather than an evaluation of method performance. A future systematic review with explicit quality criteria and standardized comparison protocols would be a useful complement to this work.

\section*{Conclusion}

The reviewed literature shows that physics-informed machine learning has made progress in seismic wave propagation modeling, both as a standalone approach and in combination with standard numerical solvers. We identified three ways of incorporating physical knowledge: observational bias, inductive bias, and learning bias. Each has shown specific advantages, and formulations that combine them are beginning to appear in the literature more frequently.

Standard numerical methods, such as finite difference and spectral element methods, remain the foundation of seismological workflows due to their accuracy and well-established theoretical basis. However, these methods face known limitations in inverse problems and high-dimensional settings, where the cost of repeated forward simulations can be high. Machine learning approaches, such as neural networks, neural operators, and PINNs, have reduced inference time in some cases by one to three orders of magnitude; however, these results generally require significant training effort and have been demonstrated mostly on synthetic, low-dimensional problems.

Some challenges remain before these methods can be used extensively in practice. Training times, generalization to three-dimensional and heterogeneous media, spectral bias toward low-frequency components, and the absence of uncertainty estimates are recurring limitations across the reviewed studies. In addition, comparisons between machine learning and numerical methods are often not conducted at equivalent accuracy levels or against suitable numerical baselines, which limits the conclusions that can be drawn about relative performance.

Future work should focus on three directions: hybrid formulations that combine standard numerical methods with machine learning components, benchmarking practices that compare methods under consistent accuracy and computational conditions, and validation using field or experimental data, which is currently limited in the literature. Addressing these points will be important for the development of machine learning methods that are both physically consistent and applicable to realistic seismological problems.

\section*{CRediT authorship contribution statement}

\textbf{Óscar Rincón Cardeño}: Writing -- original draft, Visualization, Software, Methodology, Investigation, Formal analysis, Data curation, Conceptualization. \textbf{Gregorio Pérez-Bernal}: Writing -- review \& editing, Conceptualization. \textbf{Silvana Montoya-Noguera}: Writing -- review \& editing, Supervision, Funding acquisition, Conceptualization. \textbf{Nicolás Guarín-Zapata}: Writing -- review \& editing, Visualization, Supervision, Software, Project administration, Methodology, Funding acquisition, Formal analysis, Conceptualization.

\section*{Declaration of competing interest}

The authors declare that they have no known competing financial interests or personal relationships that could have appeared to influence the work reported in this paper.

\section*{Acknowledgment}

This work was supported by Universidad EAFIT, Colombia, Program No. 12330032023.





\bibliography{sn-bibliography}


\begin{thebibliography}{99}
\ifx \bisbn   \undefined \def \bisbn  #1{ISBN #1}\fi
\ifx \binits  \undefined \def \binits#1{#1}\fi
\ifx \bauthor  \undefined \def \bauthor#1{#1}\fi
\ifx \batitle  \undefined \def \batitle#1{#1}\fi
\ifx \bjtitle  \undefined \def \bjtitle#1{#1}\fi
\ifx \bvolume  \undefined \def \bvolume#1{\textbf{#1}}\fi
\ifx \byear  \undefined \def \byear#1{#1}\fi
\ifx \bissue  \undefined \def \bissue#1{#1}\fi
\ifx \bfpage  \undefined \def \bfpage#1{#1}\fi
\ifx \blpage  \undefined \def \blpage #1{#1}\fi
\ifx \burl  \undefined \def \burl#1{\textsf{#1}}\fi
\ifx \doiurl  \undefined \def \doiurl#1{\url{https://doi.org/#1}}\fi
\ifx \betal  \undefined \def \betal{\textit{et al.}}\fi
\ifx \binstitute  \undefined \def \binstitute#1{#1}\fi
\ifx \binstitutionaled  \undefined \def \binstitutionaled#1{#1}\fi
\ifx \bctitle  \undefined \def \bctitle#1{#1}\fi
\ifx \beditor  \undefined \def \beditor#1{#1}\fi
\ifx \bpublisher  \undefined \def \bpublisher#1{#1}\fi
\ifx \bbtitle  \undefined \def \bbtitle#1{#1}\fi
\ifx \bedition  \undefined \def \bedition#1{#1}\fi
\ifx \bseriesno  \undefined \def \bseriesno#1{#1}\fi
\ifx \blocation  \undefined \def \blocation#1{#1}\fi
\ifx \bsertitle  \undefined \def \bsertitle#1{#1}\fi
\ifx \bsnm \undefined \def \bsnm#1{#1}\fi
\ifx \bsuffix \undefined \def \bsuffix#1{#1}\fi
\ifx \bparticle \undefined \def \bparticle#1{#1}\fi
\ifx \barticle \undefined \def \barticle#1{#1}\fi
\bibcommenthead
\ifx \bconfdate \undefined \def \bconfdate #1{#1}\fi
\ifx \botherref \undefined \def \botherref #1{#1}\fi
\ifx \url \undefined \def \url#1{\textsf{#1}}\fi
\ifx \bchapter \undefined \def \bchapter#1{#1}\fi
\ifx \bbook \undefined \def \bbook#1{#1}\fi
\ifx \bcomment \undefined \def \bcomment#1{#1}\fi
\ifx \oauthor \undefined \def \oauthor#1{#1}\fi
\ifx \citeauthoryear \undefined \def \citeauthoryear#1{#1}\fi
\ifx \endbibitem  \undefined \def \endbibitem {}\fi
\ifx \bconflocation  \undefined \def \bconflocation#1{#1}\fi
\ifx \arxivurl  \undefined \def \arxivurl#1{\textsf{#1}}\fi
\csname PreBibitemsHook\endcsname

\bibitem[\protect\citeauthoryear{Aminzadeh et~al.}{1994}]{aminzadeh_segeaeg_1994}
\begin{barticle}
\bauthor{\bsnm{Aminzadeh}, \binits{F.}},
\bauthor{\bsnm{Burkhard}, \binits{N.}},
\bauthor{\bsnm{Nicoletis}, \binits{L.}},
\bauthor{\bsnm{Rocca}, \binits{F.}},
\bauthor{\bsnm{Wyatt}, \binits{K.}}:
\batitle{{SEG}/{EAEG} 3-{D} modeling project: 2nd update}.
\bjtitle{The Leading Edge}
\bvolume{13}(\bissue{9}),
\bfpage{949}--\blpage{952}
(\byear{1994})
\doiurl{10.1190/1.1437054}
\end{barticle}
\endbibitem

\bibitem[\protect\citeauthoryear{Auriol et~al.}{2021}]{auriol_sensing_2021}
\begin{botherref}
\oauthor{\bsnm{Auriol}, \binits{J.}},
\oauthor{\bsnm{Kazemi}, \binits{N.}},
\oauthor{\bsnm{Niculescu}, \binits{S.-I.}}:
Sensing and computational frameworks for improving drill-string dynamics estimation.
Mechanical Systems and Signal Processing
\textbf{160}
(2021)
\doiurl{10.1016/j.ymssp.2021.107836}
\end{botherref}
\endbibitem

\bibitem[\protect\citeauthoryear{Alsadi}{2017}]{alsadi_seismic_2017}
\begin{bbook}
\bauthor{\bsnm{Alsadi}, \binits{H.N.}}:
\bbtitle{Seismic {Hydrocarbon} {Exploration}}.
\bsertitle{Advances in {Oil} and {Gas} {Exploration} \& {Production}}.
\bpublisher{Springer},
\blocation{Cham}
(\byear{2017}).
\doiurl{10.1007/978-3-319-40436-3} .
\burl{http://link.springer.com/10.1007/978-3-319-40436-3}
\end{bbook}
\endbibitem

\bibitem[\protect\citeauthoryear{Arksey and O'Malley}{2005}]{arksey_scoping_2005}
\begin{barticle}
\bauthor{\bsnm{Arksey}, \binits{H.}},
\bauthor{\bsnm{O'Malley}, \binits{L.}}:
\batitle{Scoping studies: towards a methodological framework}.
\bjtitle{International Journal of Social Research Methodology}
\bvolume{8}(\bissue{1}),
\bfpage{19}--\blpage{32}
(\byear{2005})
\doiurl{10.1080/1364557032000119616}
\end{barticle}
\endbibitem

\bibitem[\protect\citeauthoryear{Alfarhan et~al.}{2024}]{alfarhan_robust_2024}
\begin{botherref}
\oauthor{\bsnm{Alfarhan}, \binits{M.}},
\oauthor{\bsnm{Ravasi}, \binits{M.}},
\oauthor{\bsnm{Chen}, \binits{F.}},
\oauthor{\bsnm{Alkhalifah}, \binits{T.}}:
Robust {Full} {Waveform} {Inversion} with deep {Hessian} deblurring.
arXiv
(2024).
\doiurl{10.48550/arXiv.2403.17518}
\end{botherref}
\endbibitem

\bibitem[\protect\citeauthoryear{Barba}{2022}]{barba_defining_2022}
\begin{barticle}
\bauthor{\bsnm{Barba}, \binits{L.A.}}:
\batitle{Defining the {Role} of {Open} {Source} {Software} in {Research} {Reproducibility}}.
\bjtitle{Computer}
\bvolume{55}(\bissue{8}),
\bfpage{40}--\blpage{48}
(\byear{2022})
\doiurl{10.1109/MC.2022.3177133}
\end{barticle}
\endbibitem

\bibitem[\protect\citeauthoryear{Bellman et~al.}{1957}]{bellman_dynamic_1957}
\begin{bbook}
\bauthor{\bsnm{Bellman}, \binits{R.}},
\bauthor{\bsnm{Bellman}, \binits{R.E.}},
\bauthor{\bsnm{Corporation}, \binits{R.}}:
\bbtitle{Dynamic {Programming}}.
\bsertitle{Rand {Corporation} research study}.
\bpublisher{Princeton University Press}, \blocation{???}
(\byear{1957})
\end{bbook}
\endbibitem

\bibitem[\protect\citeauthoryear{Blechschmidt and Ernst}{2021}]{blechschmidt_three_2021}
\begin{barticle}
\bauthor{\bsnm{Blechschmidt}, \binits{J.}},
\bauthor{\bsnm{Ernst}, \binits{O.G.}}:
\batitle{Three ways to solve partial differential equations with neural networks — {A} review}.
\bjtitle{GAMM-Mitteilungen}
\bvolume{44}(\bissue{2}),
\bfpage{202100006}
(\byear{2021})
\doiurl{10.1002/gamm.202100006}
\end{barticle}
\endbibitem

\bibitem[\protect\citeauthoryear{Bezanson et~al.}{2014}]{bezanson_julia_2014}
\begin{botherref}
\oauthor{\bsnm{Bezanson}, \binits{J.}},
\oauthor{\bsnm{Edelman}, \binits{A.}},
\oauthor{\bsnm{Karpinski}, \binits{S.}},
\oauthor{\bsnm{Shah}, \binits{V.B.}}:
Julia: {A} {Fresh} {Approach} to {Numerical} {Computing}
(2014)
\end{botherref}
\endbibitem

\bibitem[\protect\citeauthoryear{Burden et~al.}{2016}]{burden_numerical_2016}
\begin{bbook}
\bauthor{\bsnm{Burden}, \binits{R.L.}},
\bauthor{\bsnm{Faires}, \binits{J.D.}},
\bauthor{\bsnm{Burden}, \binits{A.M.}}:
\bbtitle{Numerical {Analysis}}.
\bpublisher{Cengage Learning},
\blocation{Boston, MA}
(\byear{2016})
\end{bbook}
\endbibitem

\bibitem[\protect\citeauthoryear{Bradbury et~al.}{2018}]{jax2018github}
\begin{botherref}
\oauthor{\bsnm{Bradbury}, \binits{J.}},
\oauthor{\bsnm{Frostig}, \binits{R.}},
\oauthor{\bsnm{Hawkins}, \binits{P.}},
\oauthor{\bsnm{Johnson}, \binits{M.J.}},
\oauthor{\bsnm{Katariya}, \binits{Y.}},
\oauthor{\bsnm{Leary}, \binits{C.}},
\oauthor{\bsnm{Maclaurin}, \binits{D.}},
\oauthor{\bsnm{Necula}, \binits{G.}},
\oauthor{\bsnm{Paszke}, \binits{A.}},
\oauthor{\bsnm{Vander{P}las}, \binits{J.}},
\oauthor{\bsnm{Wanderman-{M}ilne}, \binits{S.}},
\oauthor{\bsnm{Zhang}, \binits{Q.}}:
{JAX}: composable transformations of {P}ython+{N}um{P}y programs
(2018).
\url{http://github.com/jax-ml/jax}
\end{botherref}
\endbibitem

\bibitem[\protect\citeauthoryear{Bafghi and Raissi}{2023}]{bafghi_pinns-torch_2023}
\begin{bchapter}
\bauthor{\bsnm{Bafghi}, \binits{R.A.}},
\bauthor{\bsnm{Raissi}, \binits{M.}}:
\bctitle{{PINNs}-{Torch}: {Enhancing} {Speed} and {Usability} of {Physics}-{Informed} {Neural} {Networks} with {PyTorch}}.
(\byear{2023})
\end{bchapter}
\endbibitem

\bibitem[\protect\citeauthoryear{Brandolin et~al.}{2024}]{brandolin_pinnslope_2024}
\begin{barticle}
\bauthor{\bsnm{Brandolin}, \binits{F.}},
\bauthor{\bsnm{Ravasi}, \binits{M.}},
\bauthor{\bsnm{Alkhalifah}, \binits{T.}}:
\batitle{{PINNslope}: {Seismic} data interpolation and local slope estimation with physics informed neural networks}.
\bjtitle{Geophysics}
\bvolume{89}(\bissue{4}),
\bfpage{331}--\blpage{345}
(\byear{2024})
\doiurl{10.1190/geo2023-0323.1}
\end{barticle}
\endbibitem

\bibitem[\protect\citeauthoryear{Cuomo et~al.}{2022}]{cuomo_scientific_2022}
\begin{barticle}
\bauthor{\bsnm{Cuomo}, \binits{S.}},
\bauthor{\bsnm{Di~Cola}, \binits{V.S.}},
\bauthor{\bsnm{Giampaolo}, \binits{F.}},
\bauthor{\bsnm{Rozza}, \binits{G.}},
\bauthor{\bsnm{Raissi}, \binits{M.}},
\bauthor{\bsnm{Piccialli}, \binits{F.}}:
\batitle{Scientific {Machine} {Learning} {Through} {Physics}–{Informed} {Neural} {Networks}: {Where} we are and {What}’s {Next}}.
\bjtitle{Journal of Scientific Computing}
\bvolume{92}(\bissue{3}),
\bfpage{88}
(\byear{2022})
\doiurl{10.1007/s10915-022-01939-z}
\end{barticle}
\endbibitem

\bibitem[\protect\citeauthoryear{Cao et~al.}{2020}]{cao_near-real-time_2020}
\begin{barticle}
\bauthor{\bsnm{Cao}, \binits{R.}},
\bauthor{\bsnm{Earp}, \binits{S.}},
\bauthor{\bsnm{De~Ridder}, \binits{S.A.L.}},
\bauthor{\bsnm{Curtis}, \binits{A.}},
\bauthor{\bsnm{Galetti}, \binits{E.}}:
\batitle{Near-real-time near-surface {3D} seismic velocity and uncertainty models by wavefield gradiometry and neural network inversion of ambient seismic noise}.
\bjtitle{Geophysics}
\bvolume{85}(\bissue{1}),
\bfpage{13}--\blpage{27}
(\byear{2020})
\doiurl{10.1190/geo2018-0562.1}
\end{barticle}
\endbibitem

\bibitem[\protect\citeauthoryear{Campos et~al.}{2019}]{campos_empirical_2019}
\begin{bchapter}
\bauthor{\bsnm{Campos}, \binits{L.R.}},
\bauthor{\bsnm{Nogueira}, \binits{P.}},
\bauthor{\bsnm{Moreira}, \binits{D.}},
\bauthor{\bsnm{Nascimento}, \binits{E.G.S.}}:
\bctitle{An empirical analysis of the influence of seismic data modeling for estimating velocity models with fully convolutional networks},
vol. \bseriesno{1},
pp. \bfpage{93}--\blpage{98}
(\byear{2019})
\end{bchapter}
\endbibitem

\bibitem[\protect\citeauthoryear{Chen et~al.}{2020}]{chen_neurodiffeq_2020}
\begin{barticle}
\bauthor{\bsnm{Chen}, \binits{F.}},
\bauthor{\bsnm{Sondak}, \binits{D.}},
\bauthor{\bsnm{Protopapas}, \binits{P.}},
\bauthor{\bsnm{Mattheakis}, \binits{M.}},
\bauthor{\bsnm{Liu}, \binits{S.}},
\bauthor{\bsnm{Agarwal}, \binits{D.}},
\bauthor{\bsnm{Giovanni}, \binits{M.D.}}:
\batitle{{NeuroDiffEq}: {A} {Python} package for solving differential equations with neural networks}.
\bjtitle{Journal of Open Source Software}
\bvolume{5}(\bissue{46}),
\bfpage{1931}
(\byear{2020})
\doiurl{10.21105/joss.01931}
\end{barticle}
\endbibitem

\bibitem[\protect\citeauthoryear{Das et~al.}{2018}]{das_surrogate_2018}
\begin{barticle}
\bauthor{\bsnm{Das}, \binits{S.}},
\bauthor{\bsnm{Chen}, \binits{X.}},
\bauthor{\bsnm{Hobson}, \binits{M.P.}},
\bauthor{\bsnm{Phadke}, \binits{S.}},
\bauthor{\bsnm{Beest}, \binits{B.}},
\bauthor{\bsnm{Goudswaard}, \binits{J.}},
\bauthor{\bsnm{Hohl}, \binits{D.}}:
\batitle{Surrogate regression modelling for fast seismogram generation and detection of microseismic events in heterogeneous velocity models}.
\bjtitle{Geophysical Journal International}
\bvolume{215}(\bissue{2}),
\bfpage{1257}--\blpage{1290}
(\byear{2018})
\doiurl{10.1093/GJI/GGY283}
\end{barticle}
\endbibitem

\bibitem[\protect\citeauthoryear{Deng et~al.}{2023}]{deng_physics-informed_2023}
\begin{barticle}
\bauthor{\bsnm{Deng}, \binits{W.}},
\bauthor{\bsnm{Nguyen}, \binits{K.T.P.}},
\bauthor{\bsnm{Medjaher}, \binits{K.}},
\bauthor{\bsnm{Gogu}, \binits{C.}},
\bauthor{\bsnm{Morio}, \binits{J.}}:
\batitle{Physics-informed machine learning in prognostics and health management: {State} of the art and challenges}.
\bjtitle{Applied Mathematical Modelling}
\bvolume{124},
\bfpage{325}--\blpage{352}
(\byear{2023})
\doiurl{10.1016/j.apm.2023.07.011}
\end{barticle}
\endbibitem

\bibitem[\protect\citeauthoryear{Dissanayake and Phan-Thien}{1994}]{dissanayake_neural-network-based_1994}
\begin{barticle}
\bauthor{\bsnm{Dissanayake}, \binits{M.W.M.G.}},
\bauthor{\bsnm{Phan-Thien}, \binits{N.}}:
\batitle{Neural-network-based approximations for solving partial differential equations}.
\bjtitle{Communications in Numerical Methods in Engineering}
\bvolume{10}(\bissue{3}),
\bfpage{195}--\blpage{201}
(\byear{1994})
\doiurl{10.1002/cnm.1640100303}
\end{barticle}
\endbibitem

\bibitem[\protect\citeauthoryear{Drummond}{2009}]{drummond_replicability_2009}
\begin{bchapter}
\bauthor{\bsnm{Drummond}, \binits{D.C.}}:
\bctitle{Replicability is not {Reproducibility}: {Nor} is it {Good} {Science}}.
(\byear{2009})
\end{bchapter}
\endbibitem

\bibitem[\protect\citeauthoryear{Dhara and Sen}{2022}]{dhara_elastic-adjointnet_2022}
\begin{bchapter}
\bauthor{\bsnm{Dhara}, \binits{A.}},
\bauthor{\bsnm{Sen}, \binits{M.}}:
\bctitle{Elastic-{AdjointNet}: {A} physics guided deep autoencoder to overcome cross talk effects in multiparameter full waveform inversion},
vol. \bseriesno{2022-August},
pp. \bfpage{882}--\blpage{886}
(\byear{2022}).
\doiurl{10.1190/image2022-3745050.1}
\end{bchapter}
\endbibitem

\bibitem[\protect\citeauthoryear{Escapil-Inchauspé and Ruz}{2023}]{escapil-inchauspe_hyper-parameter_2023}
\begin{botherref}
\oauthor{\bsnm{Escapil-Inchauspé}, \binits{P.}},
\oauthor{\bsnm{Ruz}, \binits{G.A.}}:
Hyper-parameter tuning of physics-informed neural networks: {Application} to {Helmholtz} problems.
arXiv.
arXiv:2205.06704 [math]
(2023).
\doiurl{10.48550/arXiv.2205.06704}
\end{botherref}
\endbibitem

\bibitem[\protect\citeauthoryear{Fang et~al.}{2024}]{fang_shear-wave_2024}
\begin{botherref}
\oauthor{\bsnm{Fang}, \binits{Z.}},
\oauthor{\bsnm{Ba}, \binits{J.}},
\oauthor{\bsnm{Guo}, \binits{Q.}},
\oauthor{\bsnm{Xiong}, \binits{F.}}:
Shear-wave velocity prediction of tight reservoirs based on poroelasticity theory: {A} comparative study of deep neural network and rock physics model.
Geoenergy Science and Engineering
\textbf{240}
(2024)
\doiurl{10.1016/j.geoen.2024.213028}
\end{botherref}
\endbibitem

\bibitem[\protect\citeauthoryear{Feng et~al.}{2022}]{feng_intriguing_2022}
\begin{barticle}
\bauthor{\bsnm{Feng}, \binits{Y.}},
\bauthor{\bsnm{Chen}, \binits{Y.}},
\bauthor{\bsnm{Feng}, \binits{S.}},
\bauthor{\bsnm{Jin}, \binits{P.}},
\bauthor{\bsnm{Liu}, \binits{Z.}},
\bauthor{\bsnm{Lin}, \binits{Y.}}:
\batitle{An {Intriguing} {Property} of {Geophysics} {Inversion}}.
\bjtitle{arXiv (Cornell University)}
(\byear{2022})
\doiurl{10.48550/arxiv.2204.13731}
\end{barticle}
\endbibitem

\bibitem[\protect\citeauthoryear{Fu et~al.}{2019}]{fu_seismic_2019}
\begin{bchapter}
\bauthor{\bsnm{Fu}, \binits{H.}},
\bauthor{\bsnm{Zhang}, \binits{Y.}},
\bauthor{\bsnm{Ma}, \binits{M.}}:
\bctitle{Seismic waveform inversion using a neural network-based forward},
vol. \bseriesno{1324}
(\byear{2019}).
\doiurl{10.1088/1742-6596/1324/1/012043}
\end{bchapter}
\endbibitem

\bibitem[\protect\citeauthoryear{Gelboim et~al.}{2023}]{gelboim_encoder-decoder_2023}
\begin{botherref}
\oauthor{\bsnm{Gelboim}, \binits{M.}},
\oauthor{\bsnm{Adler}, \binits{A.}},
\oauthor{\bsnm{Sun}, \binits{Y.}},
\oauthor{\bsnm{Araya-Polo}, \binits{M.}}:
Encoder-{Decoder} {Architecture} for {3D} {Seismic} {Inversion}.
Sensors
\textbf{23}(1)
(2023)
\doiurl{10.3390/s23010061}
\end{botherref}
\endbibitem

\bibitem[\protect\citeauthoryear{Goodfellow et~al.}{2016}]{goodfellow_deep_2016}
\begin{bbook}
\bauthor{\bsnm{Goodfellow}, \binits{I.}},
\bauthor{\bsnm{Bengio}, \binits{Y.}},
\bauthor{\bsnm{Courville}, \binits{A.}}:
\bbtitle{Deep {Learning}}.
\bpublisher{MIT Press},
\blocation{Cambridge, MA}
(\byear{2016})
\end{bbook}
\endbibitem

\bibitem[\protect\citeauthoryear{Gao et~al.}{2024}]{gao_underground_2024}
\begin{barticle}
\bauthor{\bsnm{Gao}, \binits{K.}},
\bauthor{\bsnm{Creasy}, \binits{N.M.}},
\bauthor{\bsnm{Huang}, \binits{L.}},
\bauthor{\bsnm{Gross}, \binits{M.R.}}:
\batitle{Underground hydrogen storage leakage detection and characterization based on machine learning of sparse seismic data}.
\bjtitle{International Journal of Hydrogen Energy}
\bvolume{61},
\bfpage{137}--\blpage{161}
(\byear{2024})
\doiurl{10.1016/j.ijhydene.2024.02.296}
\end{barticle}
\endbibitem

\bibitem[\protect\citeauthoryear{Gosselin et~al.}{2022}]{gosselin_review_2022}
\begin{barticle}
\bauthor{\bsnm{Gosselin}, \binits{J.M.}},
\bauthor{\bsnm{Dosso}, \binits{S.E.}},
\bauthor{\bsnm{Askan}, \binits{A.}},
\bauthor{\bsnm{Wathelet}, \binits{M.}},
\bauthor{\bsnm{Savvaidis}, \binits{A.}},
\bauthor{\bsnm{Cassidy}, \binits{J.F.}}:
\batitle{A review of inverse methods in seismic site characterization}.
\bjtitle{Journal of Seismology}
\bvolume{26}(\bissue{4}),
\bfpage{781}--\blpage{821}
(\byear{2022})
\doiurl{10.1007/s10950-021-10047-8}
\end{barticle}
\endbibitem

\bibitem[\protect\citeauthoryear{Grossmann et~al.}{2024}]{grossmann_can_2024}
\begin{barticle}
\bauthor{\bsnm{Grossmann}, \binits{T.G.}},
\bauthor{\bsnm{Komorowska}, \binits{U.J.}},
\bauthor{\bsnm{Latz}, \binits{J.}},
\bauthor{\bsnm{Schönlieb}, \binits{C.-B.}}:
\batitle{Can physics-informed neural networks beat the finite element method?}
\bjtitle{IMA Journal of Applied Mathematics}
\bvolume{89}(\bissue{1}),
\bfpage{143}--\blpage{174}
(\byear{2024})
\doiurl{10.1093/imamat/hxae011}
\end{barticle}
\endbibitem

\bibitem[\protect\citeauthoryear{Guo et~al.}{2024}]{guo_parametric_2024}
\begin{barticle}
\bauthor{\bsnm{Guo}, \binits{K.}},
\bauthor{\bsnm{Zong}, \binits{Z.}},
\bauthor{\bsnm{Yang}, \binits{J.}},
\bauthor{\bsnm{Tan}, \binits{Y.}}:
\batitle{Parametric elastic full waveform inversion with convolutional neural network}.
\bjtitle{Acta Geophysica}
\bvolume{72}(\bissue{2}),
\bfpage{673}--\blpage{687}
(\byear{2024})
\doiurl{10.1007/s11600-023-01123-3}
\end{barticle}
\endbibitem

\bibitem[\protect\citeauthoryear{Huang et~al.}{2022}]{huang_applying_2022}
\begin{bchapter}
\bauthor{\bsnm{Huang}, \binits{L.}},
\bauthor{\bsnm{Clee}, \binits{E.}},
\bauthor{\bsnm{Ranasinghe}, \binits{N.}}:
\bctitle{Applying scientific machine learning to improve seismic wave simulation and inversion}.
In: \bbtitle{Advances in {Subsurface} {Data} {Analytics}: {Traditional} and {Physics}-{Based} {Machine} {Learning}},
pp. \bfpage{167}--\blpage{192}
(\byear{2022})
\end{bchapter}
\endbibitem

\bibitem[\protect\citeauthoryear{Hao et~al.}{2023}]{hao_physics-informed_2023}
\begin{botherref}
\oauthor{\bsnm{Hao}, \binits{Z.}},
\oauthor{\bsnm{Liu}, \binits{S.}},
\oauthor{\bsnm{Zhang}, \binits{Y.}},
\oauthor{\bsnm{Ying}, \binits{C.}},
\oauthor{\bsnm{Feng}, \binits{Y.}},
\oauthor{\bsnm{Su}, \binits{H.}},
\oauthor{\bsnm{Zhu}, \binits{J.}}:
Physics-{Informed} {Machine} {Learning}: {A} {Survey} on {Problems}, {Methods} and {Applications}.
arXiv.
arXiv:2211.08064 [cs, math]
(2023).
\url{http://arxiv.org/abs/2211.08064}
\end{botherref}
\endbibitem

\bibitem[\protect\citeauthoryear{Hornik}{1991}]{hornik_approximation_1991}
\begin{barticle}
\bauthor{\bsnm{Hornik}, \binits{K.}}:
\batitle{Approximation capabilities of multilayer feedforward networks}.
\bjtitle{Neural Networks}
\bvolume{4}(\bissue{2}),
\bfpage{251}--\blpage{257}
(\byear{1991})
\doiurl{10.1016/0893-6080(91)90009-T}
\end{barticle}
\endbibitem

\bibitem[\protect\citeauthoryear{Haghighat et~al.}{2021}]{haghighat_physics-informed_2021}
\begin{barticle}
\bauthor{\bsnm{Haghighat}, \binits{E.}},
\bauthor{\bsnm{Raissi}, \binits{M.}},
\bauthor{\bsnm{Moure}, \binits{A.}},
\bauthor{\bsnm{Gomez}, \binits{H.}},
\bauthor{\bsnm{Juanes}, \binits{R.}}:
\batitle{A physics-informed deep learning framework for inversion and surrogate modeling in solid mechanics}.
\bjtitle{Computer Methods in Applied Mechanics and Engineering}
\bvolume{379},
\bfpage{113741}
(\byear{2021})
\doiurl{10.1016/j.cma.2021.113741}
\end{barticle}
\endbibitem

\bibitem[\protect\citeauthoryear{Hateley et~al.}{2020a}]{hateley_deep_2020}
\begin{barticle}
\bauthor{\bsnm{Hateley}, \binits{J.C.}},
\bauthor{\bsnm{Roberts}, \binits{J.}},
\bauthor{\bsnm{Mylonakis}, \binits{K.}},
\bauthor{\bsnm{Yang}, \binits{X.}}:
\batitle{Deep learning seismic substructure detection using the {Frozen} {Gaussian} approximation}.
\bjtitle{Journal of Computational Physics}
\bvolume{409},
\bfpage{109313}
(\byear{2020})
\doiurl{10.1016/j.jcp.2020.109313}
\end{barticle}
\endbibitem

\bibitem[\protect\citeauthoryear{Hateley et~al.}{2020b}]{hateleyDeepLearningSeismic2020}
\begin{barticle}
\bauthor{\bsnm{Hateley}, \binits{J.C.}},
\bauthor{\bsnm{Roberts}, \binits{J.}},
\bauthor{\bsnm{Mylonakis}, \binits{K.}},
\bauthor{\bsnm{Yang}, \binits{X.}}:
\batitle{Deep learning seismic substructure detection using the {Frozen} {Gaussian} approximation}.
\bjtitle{Journal of Computational Physics}
\bvolume{409},
\bfpage{109313}
(\byear{2020})
\doiurl{10.1016/j.jcp.2020.109313}
\end{barticle}
\endbibitem

\bibitem[\protect\citeauthoryear{Igel}{2017}]{igel_computational_2017}
\begin{bbook}
\bauthor{\bsnm{Igel}, \binits{H.}}:
\bbtitle{Computational Seismology: a Practical Introduction}.
\bpublisher{Oxford University Press},
\blocation{Oxford}
(\byear{2017})
\end{bbook}
\endbibitem

\bibitem[\protect\citeauthoryear{JingBo et~al.}{2023}]{jingbo_research_2023}
\begin{barticle}
\bauthor{\bsnm{JingBo}, \binits{Z.O.U.}},
\bauthor{\bsnm{Cai}, \binits{L.I.U.}},
\bauthor{\bsnm{PengFei}, \binits{Z.}}:
\batitle{Research progress of physics-informed neural network in seismic wave modeling}.
\bjtitle{Progress in Geophysics}
\bvolume{38}(\bissue{1}),
\bfpage{430}--\blpage{448}
(\byear{2023})
\doiurl{10.6038/pg2023GG0142}
\end{barticle}
\endbibitem

\bibitem[\protect\citeauthoryear{Ji et~al.}{2024}]{ji_efficient_2024}
\begin{barticle}
\bauthor{\bsnm{Ji}, \binits{D.}},
\bauthor{\bsnm{Li}, \binits{C.}},
\bauthor{\bsnm{Zhai}, \binits{C.}},
\bauthor{\bsnm{Cao}, \binits{Z.}}:
\batitle{An {Efficient} {Platform} for {Numerical} {Modeling} of {Partial} {Differential} {Equations}}.
\bjtitle{IEEE Transactions on Geoscience and Remote Sensing}
\bvolume{62},
\bfpage{1}--\blpage{13}
(\byear{2024})
\doiurl{10.1109/TGRS.2024.3409620}
\end{barticle}
\endbibitem

\bibitem[\protect\citeauthoryear{Karniadakis et~al.}{2021}]{karniadakis_physics-informed_2021}
\begin{barticle}
\bauthor{\bsnm{Karniadakis}, \binits{G.E.}},
\bauthor{\bsnm{Kevrekidis}, \binits{I.G.}},
\bauthor{\bsnm{Lu}, \binits{L.}},
\bauthor{\bsnm{Perdikaris}, \binits{P.}},
\bauthor{\bsnm{Wang}, \binits{S.}},
\bauthor{\bsnm{Yang}, \binits{L.}}:
\batitle{Physics-informed machine learning}.
\bjtitle{Nature Reviews Physics}
\bvolume{3}(\bissue{6}),
\bfpage{422}--\blpage{440}
(\byear{2021})
\doiurl{10.1038/s42254-021-00314-5} .
\bcomment{Number: 6}
\end{barticle}
\endbibitem

\bibitem[\protect\citeauthoryear{Komatitsch and Martin}{2007}]{komatitsch_unsplit_2007}
\begin{botherref}
\oauthor{\bsnm{Komatitsch}, \binits{D.}},
\oauthor{\bsnm{Martin}, \binits{R.}}:
An unsplit convolutional {Perfectly} {Matched} {Layer} improved at grazing incidence for the seismic wave equation.
Geophysics
\textbf{72}
(2007)
\doiurl{10.1190/1.2757586}
\end{botherref}
\endbibitem

\bibitem[\protect\citeauthoryear{Karimpouli and Tahmasebi}{2020}]{karimpouli_physics_2020}
\begin{barticle}
\bauthor{\bsnm{Karimpouli}, \binits{S.}},
\bauthor{\bsnm{Tahmasebi}, \binits{P.}}:
\batitle{Physics informed machine learning: {Seismic} wave equation}.
\bjtitle{Geoscience Frontiers}
\bvolume{11}(\bissue{6}),
\bfpage{1993}--\blpage{2001}
(\byear{2020})
\doiurl{10.1016/j.gsf.2020.07.007}
\end{barticle}
\endbibitem

\bibitem[\protect\citeauthoryear{Komatitsch et~al.}{2023}]{komatitsch_specfemspecfem2d_2023}
\begin{botherref}
\oauthor{\bsnm{Komatitsch}, \binits{D.}},
\oauthor{\bsnm{Tromp}, \binits{J.}},
\oauthor{\bsnm{Gharti}, \binits{H.N.}},
\oauthor{\bsnm{Peter}, \binits{D.}},
\oauthor{\bsnm{Cano}, \binits{E.V.}},
\oauthor{\bsnm{Bachmann}, \binits{E.}},
\oauthor{\bsnm{Bottero}, \binits{A.}},
\oauthor{\bsnm{Brissaud}, \binits{Q.}},
\oauthor{\bsnm{Chow}, \binits{B.}},
\oauthor{\bsnm{Cristini}, \binits{P.}},
\oauthor{\bsnm{Cui}, \binits{C.}},
\oauthor{\bsnm{Gassmoeller}, \binits{R.}},
\oauthor{\bsnm{Gineste}, \binits{M.}},
\oauthor{\bsnm{Halpaap}, \binits{F.}},
\oauthor{\bsnm{Heien}, \binits{E.}},
\oauthor{\bsnm{Labarta}, \binits{J.}},
\oauthor{\bsnm{Lefebvre}, \binits{M.}},
\oauthor{\bsnm{Goff}, \binits{N.L.}},
\oauthor{\bsnm{Loher}, \binits{P.L.}},
\oauthor{\bsnm{Liu}, \binits{Q.}},
\oauthor{\bsnm{Liu}, \binits{Q.}},
\oauthor{\bsnm{Liu}, \binits{Y.}},
\oauthor{\bsnm{Liu}, \binits{Z.}},
\oauthor{\bsnm{Luet}, \binits{D.}},
\oauthor{\bsnm{Martin}, \binits{R.}},
\oauthor{\bsnm{Matzen}, \binits{R.}},
\oauthor{\bsnm{Modrak}, \binits{R.}},
\oauthor{\bsnm{Morency}, \binits{C.}},
\oauthor{\bsnm{Nagaso}, \binits{M.}},
\oauthor{\bsnm{Rosenkrantz}, \binits{E.}},
\oauthor{\bsnm{Rusmanugroho}, \binits{H.}},
\oauthor{\bsnm{Andrade}, \binits{E.S.d.}},
\oauthor{\bsnm{Tape}, \binits{C.}},
\oauthor{\bsnm{Vilotte}, \binits{J.-P.}},
\oauthor{\bsnm{Xie}, \binits{Z.}},
\oauthor{\bsnm{Zhang}, \binits{Z.}}:
{SPECFEM}/specfem2d: {SPECFEM2D} v8.1.0.
Zenodo
(2023).
\doiurl{10.5281/zenodo.10415228}
\end{botherref}
\endbibitem

\bibitem[\protect\citeauthoryear{Komatitsch et~al.}{2024}]{komatitsch_specfemspecfem3d_2024}
\begin{botherref}
\oauthor{\bsnm{Komatitsch}, \binits{D.}},
\oauthor{\bsnm{Tromp}, \binits{J.}},
\oauthor{\bsnm{Garg}, \binits{R.}},
\oauthor{\bsnm{Gharti}, \binits{H.N.}},
\oauthor{\bsnm{Nagaso}, \binits{M.}},
\oauthor{\bsnm{Oral}, \binits{E.}},
\oauthor{\bsnm{Peter}, \binits{D.}},
\oauthor{\bsnm{Afanasiev}, \binits{M.}},
\oauthor{\bsnm{Almada}, \binits{R.}},
\oauthor{\bsnm{Ampuero}, \binits{J.-P.}},
\oauthor{\bsnm{Bachmann}, \binits{E.}},
\oauthor{\bsnm{Bai}, \binits{K.}},
\oauthor{\bsnm{Basini}, \binits{P.}},
\oauthor{\bsnm{Beller}, \binits{S.}},
\oauthor{\bsnm{Bishop}, \binits{J.}},
\oauthor{\bsnm{Bissey}, \binits{F.}},
\oauthor{\bsnm{Blitz}, \binits{C.}},
\oauthor{\bsnm{Bottero}, \binits{A.}},
\oauthor{\bsnm{Bozdag}, \binits{E.}},
\oauthor{\bsnm{Casarotti}, \binits{E.}},
\oauthor{\bsnm{Charles}, \binits{J.}},
\oauthor{\bsnm{Chen}, \binits{M.}},
\oauthor{\bsnm{Cristini}, \binits{P.}},
\oauthor{\bsnm{Durochat}, \binits{C.}},
\oauthor{\bsnm{Galvez~Barron}, \binits{P.}},
\oauthor{\bsnm{Gassmoeller}, \binits{R.}},
\oauthor{\bsnm{Goeddeke}, \binits{D.}},
\oauthor{\bsnm{Grinberg}, \binits{L.}},
\oauthor{\bsnm{Gupta}, \binits{A.}},
\oauthor{\bsnm{Heien}, \binits{E.}},
\oauthor{\bsnm{Hjoerleifsdottir}, \binits{V.}},
\oauthor{\bsnm{Karakostas}, \binits{F.}},
\oauthor{\bsnm{Kientz}, \binits{S.}},
\oauthor{\bsnm{Labarta}, \binits{J.}},
\oauthor{\bsnm{Le~Goff}, \binits{N.}},
\oauthor{\bsnm{Le~Loher}, \binits{P.}},
\oauthor{\bsnm{Lefebvre}, \binits{M.}},
\oauthor{\bsnm{Liu}, \binits{Q.}},
\oauthor{\bsnm{Liu}, \binits{Y.}},
\oauthor{\bsnm{Luet}, \binits{D.}},
\oauthor{\bsnm{Luo}, \binits{Y.}},
\oauthor{\bsnm{Maggi}, \binits{A.}},
\oauthor{\bsnm{Magnoni}, \binits{F.}},
\oauthor{\bsnm{Martin}, \binits{R.}},
\oauthor{\bsnm{Matzen}, \binits{R.}},
\oauthor{\bsnm{McBain}, \binits{G.D.}},
\oauthor{\bsnm{McRitchie}, \binits{D.}},
\oauthor{\bsnm{Meschede}, \binits{M.}},
\oauthor{\bsnm{Messmer}, \binits{P.}},
\oauthor{\bsnm{Michea}, \binits{D.}},
\oauthor{\bsnm{Miller}, \binits{D.}},
\oauthor{\bsnm{Modrak}, \binits{R.}},
\oauthor{\bsnm{Monteiller}, \binits{V.}},
\oauthor{\bsnm{Morency}, \binits{C.}},
\oauthor{\bsnm{Nadh~Somala}, \binits{S.}},
\oauthor{\bsnm{Nissen-Meyer}, \binits{T.}},
\oauthor{\bsnm{Pouget}, \binits{K.}},
\oauthor{\bsnm{Rietmann}, \binits{M.}},
\oauthor{\bsnm{Andrade}, \binits{E.}},
\oauthor{\bsnm{Savage}, \binits{B.}},
\oauthor{\bsnm{Schuberth}, \binits{B.}},
\oauthor{\bsnm{Sieminski}, \binits{A.}},
\oauthor{\bsnm{Smith}, \binits{J.}},
\oauthor{\bsnm{Strand}, \binits{L.}},
\oauthor{\bsnm{Tape}, \binits{C.}},
\oauthor{\bsnm{Valero~Cano}, \binits{E.}},
\oauthor{\bsnm{Videau}, \binits{B.}},
\oauthor{\bsnm{Vilotte}, \binits{J.-P.}},
\oauthor{\bsnm{Weng}, \binits{H.}},
\oauthor{\bsnm{Xie}, \binits{Z.}},
\oauthor{\bsnm{Zhang}, \binits{C.-H.}},
\oauthor{\bsnm{Zhu}, \binits{H.}}:
{SPECFEM}/specfem3d: {SPECFEM3D} v4.1.1.
Zenodo
(2024).
\doiurl{10.5281/zenodo.10823181}
\end{botherref}
\endbibitem

\bibitem[\protect\citeauthoryear{Kharazmi et~al.}{2019}]{kharazmi_variational_2019}
\begin{botherref}
\oauthor{\bsnm{Kharazmi}, \binits{E.}},
\oauthor{\bsnm{Zhang}, \binits{Z.}},
\oauthor{\bsnm{Karniadakis}, \binits{G.E.}}:
Variational {Physics}-{Informed} {Neural} {Networks} {For} {Solving} {Partial} {Differential} {Equations}.
arXiv
(2019).
\doiurl{10.48550/arXiv.1912.00873}
\end{botherref}
\endbibitem

\bibitem[\protect\citeauthoryear{LeCun et~al.}{2015}]{lecun_deep_2015}
\begin{barticle}
\bauthor{\bsnm{LeCun}, \binits{Y.}},
\bauthor{\bsnm{Bengio}, \binits{Y.}},
\bauthor{\bsnm{Hinton}, \binits{G.}}:
\batitle{Deep learning}.
\bjtitle{Nature}
\bvolume{521}(\bissue{7553}),
\bfpage{436}--\blpage{444}
(\byear{2015})
\doiurl{10.1038/nature14539}
\end{barticle}
\endbibitem

\bibitem[\protect\citeauthoryear{Lino et~al.}{2023}]{lino_current_2023}
\begin{barticle}
\bauthor{\bsnm{Lino}, \binits{M.}},
\bauthor{\bsnm{Fotiadis}, \binits{S.}},
\bauthor{\bsnm{Bharath}, \binits{A.A.}},
\bauthor{\bsnm{Cantwell}, \binits{C.D.}}:
\batitle{Current and emerging deep-learning methods for the simulation of fluid dynamics}.
\bjtitle{Proceedings of the Royal Society A: Mathematical, Physical and Engineering Sciences}
\bvolume{479}(\bissue{2275}),
\bfpage{20230058}
(\byear{2023})
\doiurl{10.1098/rspa.2023.0058}
\end{barticle}
\endbibitem

\bibitem[\protect\citeauthoryear{Lehmann et~al.}{2024}]{lehmann_fourier_2024}
\begin{barticle}
\bauthor{\bsnm{Lehmann}, \binits{F.}},
\bauthor{\bsnm{Gatti}, \binits{F.}},
\bauthor{\bsnm{Bertin}, \binits{M.}},
\bauthor{\bsnm{Clouteau}, \binits{D.}}:
\batitle{Fourier {Neural} {Operator} {Surrogate} {Model} to {Predict} {3D} {Seismic} {Waves} {Propagation}}.
\bjtitle{Computer Methods in Applied Mechanics and Engineering}
\bvolume{420},
\bfpage{116718}
(\byear{2024})
\doiurl{10.1016/j.cma.2023.116718}
\end{barticle}
\endbibitem

\bibitem[\protect\citeauthoryear{Lehmann et~al.}{2024}]{lehmann_multiple-input_2024}
\begin{botherref}
\oauthor{\bsnm{Lehmann}, \binits{F.}},
\oauthor{\bsnm{Gatti}, \binits{F.}},
\oauthor{\bsnm{Clouteau}, \binits{D.}}:
Multiple-{Input} {Fourier} {Neural} {Operator} ({MIFNO}) for source-dependent {3D} elastodynamics.
arXiv
(2024).
\doiurl{10.48550/arXiv.2404.10115}
\end{botherref}
\endbibitem

\bibitem[\protect\citeauthoryear{Li et~al.}{2020}]{li_neural_2020}
\begin{botherref}
\oauthor{\bsnm{Li}, \binits{Z.}},
\oauthor{\bsnm{Kovachki}, \binits{N.}},
\oauthor{\bsnm{Azizzadenesheli}, \binits{K.}},
\oauthor{\bsnm{Liu}, \binits{B.}},
\oauthor{\bsnm{Bhattacharya}, \binits{K.}},
\oauthor{\bsnm{Stuart}, \binits{A.}},
\oauthor{\bsnm{Anandkumar}, \binits{A.}}:
Neural {Operator}: {Graph} {Kernel} {Network} for {Partial} {Differential} {Equations}.
arXiv.
arXiv:2003.03485 [cs, math, stat]
(2020).
\doiurl{10.48550/arXiv.2003.03485}
\end{botherref}
\endbibitem

\bibitem[\protect\citeauthoryear{Li et~al.}{2021}]{li_fourier_2021}
\begin{botherref}
\oauthor{\bsnm{Li}, \binits{Z.}},
\oauthor{\bsnm{Kovachki}, \binits{N.}},
\oauthor{\bsnm{Azizzadenesheli}, \binits{K.}},
\oauthor{\bsnm{Liu}, \binits{B.}},
\oauthor{\bsnm{Bhattacharya}, \binits{K.}},
\oauthor{\bsnm{Stuart}, \binits{A.}},
\oauthor{\bsnm{Anandkumar}, \binits{A.}}:
Fourier {Neural} {Operator} for {Parametric} {Partial} {Differential} {Equations}.
arXiv
(2021).
\doiurl{10.48550/arXiv.2010.08895}
\end{botherref}
\endbibitem

\bibitem[\protect\citeauthoryear{Lagaris et~al.}{1998}]{lagaris_artificial_1998}
\begin{barticle}
\bauthor{\bsnm{Lagaris}, \binits{I.E.}},
\bauthor{\bsnm{Likas}, \binits{A.}},
\bauthor{\bsnm{Fotiadis}, \binits{D.I.}}:
\batitle{Artificial neural networks for solving ordinary and partial differential equations}.
\bjtitle{IEEE Transactions on Neural Networks}
\bvolume{9}(\bissue{5}),
\bfpage{987}--\blpage{1000}
(\byear{1998})
\doiurl{10.1109/72.712178} .
\bcomment{Conference Name: IEEE Transactions on Neural Networks}
\end{barticle}
\endbibitem

\bibitem[\protect\citeauthoryear{Louboutin et~al.}{2019}]{louboutin_devito_2019}
\begin{barticle}
\bauthor{\bsnm{Louboutin}, \binits{M.}},
\bauthor{\bsnm{Lange}, \binits{M.}},
\bauthor{\bsnm{Luporini}, \binits{F.}},
\bauthor{\bsnm{Kukreja}, \binits{N.}},
\bauthor{\bsnm{Witte}, \binits{P.A.}},
\bauthor{\bsnm{Herrmann}, \binits{F.J.}},
\bauthor{\bsnm{Velesko}, \binits{P.}},
\bauthor{\bsnm{Gorman}, \binits{G.J.}}:
\batitle{Devito (v3.1.0): an embedded domain-specific language for finite differences and geophysical exploration}.
\bjtitle{Geoscientific Model Development}
\bvolume{12}(\bissue{3}),
\bfpage{1165}--\blpage{1187}
(\byear{2019})
\doiurl{10.5194/gmd-12-1165-2019}
\end{barticle}
\endbibitem

\bibitem[\protect\citeauthoryear{Lu et~al.}{2021}]{lu_deepxde_2021}
\begin{barticle}
\bauthor{\bsnm{Lu}, \binits{L.}},
\bauthor{\bsnm{Meng}, \binits{X.}},
\bauthor{\bsnm{Mao}, \binits{Z.}},
\bauthor{\bsnm{Karniadakis}, \binits{G.E.}}:
\batitle{{DeepXDE}: {A} {Deep} {Learning} {Library} for {Solving} {Differential} {Equations}}.
\bjtitle{SIAM Review}
\bvolume{63}(\bissue{1}),
\bfpage{208}--\blpage{228}
(\byear{2021})
\doiurl{10.1137/19M1274067}
\end{barticle}
\endbibitem

\bibitem[\protect\citeauthoryear{Lähivaara et~al.}{2022}]{lahivaara_deep_2022}
\begin{bchapter}
\bauthor{\bsnm{Lähivaara}, \binits{T.}},
\bauthor{\bsnm{Malehmir}, \binits{A.}},
\bauthor{\bsnm{Pasanen}, \binits{A.}},
\bauthor{\bsnm{Kärkkäinen}, \binits{L.}},
\bauthor{\bsnm{Huttunen}, \binits{J.M.J.}},
\bauthor{\bsnm{Hesthaven}, \binits{J.S.}}:
\bctitle{Deep learning-based groundwater storage estimation from seismic data}.
(\byear{2022}).
\doiurl{10.3997/2214-4609.202229007}
\end{bchapter}
\endbibitem

\bibitem[\protect\citeauthoryear{Lu and Zhang}{2023}]{lu_seismic_2023}
\begin{botherref}
\oauthor{\bsnm{Lu}, \binits{C.}},
\oauthor{\bsnm{Zhang}, \binits{C.}}:
Seismic {Velocity} {Inversion} via {Physical} {Embedding} {Recurrent} {Neural} {Networks} ({RNN}).
Applied Sciences (Switzerland)
\textbf{13}(24)
(2023)
\doiurl{10.3390/app132413312}
\end{botherref}
\endbibitem

\bibitem[\protect\citeauthoryear{Mosser et~al.}{2020}]{mosser_stochastic_2020}
\begin{barticle}
\bauthor{\bsnm{Mosser}, \binits{L.}},
\bauthor{\bsnm{Dubrule}, \binits{O.}},
\bauthor{\bsnm{Blunt}, \binits{M.J.}}:
\batitle{Stochastic {Seismic} {Waveform} {Inversion} {Using} {Generative} {Adversarial} {Networks} as a {Geological} {Prior}}.
\bjtitle{Mathematical Geosciences}
\bvolume{52}(\bissue{1}),
\bfpage{53}--\blpage{79}
(\byear{2020})
\doiurl{10.1007/s11004-019-09832-6}
\end{barticle}
\endbibitem

\bibitem[\protect\citeauthoryear{Mehrkanoon et~al.}{2012}]{mehrkanoon_approximate_2012}
\begin{barticle}
\bauthor{\bsnm{Mehrkanoon}, \binits{S.}},
\bauthor{\bsnm{Falck}, \binits{T.}},
\bauthor{\bsnm{Suykens}, \binits{J.A.K.}}:
\batitle{Approximate {Solutions} to {Ordinary} {Differential} {Equations} {Using} {Least} {Squares} {Support} {Vector} {Machines}}.
\bjtitle{IEEE Transactions on Neural Networks and Learning Systems}
\bvolume{23}(\bissue{9}),
\bfpage{1356}--\blpage{1367}
(\byear{2012})
\doiurl{10.1109/TNNLS.2012.2202126} .
\bcomment{Conference Name: IEEE Transactions on Neural Networks and Learning Systems}
\end{barticle}
\endbibitem

\bibitem[\protect\citeauthoryear{McGreivy and Hakim}{2024}]{mcgreivy_weak_2024}
\begin{barticle}
\bauthor{\bsnm{McGreivy}, \binits{N.}},
\bauthor{\bsnm{Hakim}, \binits{A.}}:
\batitle{Weak baselines and reporting biases lead to overoptimism in machine learning for fluid-related partial differential equations}.
\bjtitle{Nature Machine Intelligence}
\bvolume{6}(\bissue{10}),
\bfpage{1256}--\blpage{1269}
(\byear{2024})
\doiurl{10.1038/s42256-024-00897-5}
\end{barticle}
\endbibitem

\bibitem[\protect\citeauthoryear{Moseley et~al.}{2020}]{moseley_deep_2020}
\begin{barticle}
\bauthor{\bsnm{Moseley}, \binits{B.}},
\bauthor{\bsnm{Nissen-Meyer}, \binits{T.}},
\bauthor{\bsnm{Markham}, \binits{A.}}:
\batitle{Deep learning for fast simulation of seismic waves in complex media}.
\bjtitle{Solid Earth}
\bvolume{11}(\bissue{4}),
\bfpage{1527}--\blpage{1549}
(\byear{2020})
\doiurl{10.5194/se-11-1527-2020}
\end{barticle}
\endbibitem

\bibitem[\protect\citeauthoryear{Moczo et~al.}{2007}]{moczo_finite-difference_2007}
\begin{barticle}
\bauthor{\bsnm{Moczo}, \binits{P.}},
\bauthor{\bsnm{Robertsson}, \binits{J.O.A.}},
\bauthor{\bsnm{Eisner}, \binits{L.}}:
\batitle{The {Finite}-{Difference} {Time}-{Domain} {Method} for {Modeling} of {Seismic} {Wave} {Propagation}}.
\bjtitle{Advances in Geophysics}
\bvolume{48},
\bfpage{421}--\blpage{516}
(\byear{2007})
\doiurl{10.1016/S0065-2687(06)48008-0}
\end{barticle}
\endbibitem

\bibitem[\protect\citeauthoryear{Mehrkanoon and Suykens}{2015}]{mehrkanoon_learning_2015}
\begin{barticle}
\bauthor{\bsnm{Mehrkanoon}, \binits{S.}},
\bauthor{\bsnm{Suykens}, \binits{J.A.K.}}:
\batitle{Learning solutions to partial differential equations using {LS}-{SVM}}.
\bjtitle{Neurocomputing}
\bvolume{159},
\bfpage{105}--\blpage{116}
(\byear{2015})
\doiurl{10.1016/j.neucom.2015.02.013}
\end{barticle}
\endbibitem

\bibitem[\protect\citeauthoryear{Perkel}{2020}]{perkel_challenge_2020}
\begin{barticle}
\bauthor{\bsnm{Perkel}, \binits{J.M.}}:
\batitle{Challenge to scientists: does your ten-year-old code still run?}
\bjtitle{Nature}
\bvolume{584}(\bissue{7822}),
\bfpage{656}--\blpage{658}
(\byear{2020})
\doiurl{10.1038/d41586-020-02462-7}
\end{barticle}
\endbibitem

\bibitem[\protect\citeauthoryear{Park et~al.}{2024}]{park_low-frequency_2024}
\begin{barticle}
\bauthor{\bsnm{Park}, \binits{Y.}},
\bauthor{\bsnm{Moon}, \binits{H.-J.}},
\bauthor{\bsnm{Pyun}, \binits{S.}}:
\batitle{Low-frequency marine seismic data reconstruction based on the far-field signature using a modified {U}-{Net}}.
\bjtitle{Exploration Geophysics}
\bvolume{55}(\bissue{3}),
\bfpage{263}--\blpage{276}
(\byear{2024})
\doiurl{10.1080/08123985.2024.2317129}
\end{barticle}
\endbibitem

\bibitem[\protect\citeauthoryear{Rincón-Cardeño et~al.}{2026}]{cardenoBenchmarkingPhysicsinformedNeural2026}
\begin{barticle}
\bauthor{\bsnm{Rincón-Cardeño}, \binits{O.A.}},
\bauthor{\bsnm{Perez-Bernal}, \binits{G.}},
\bauthor{\bsnm{Montoya-Noguera}, \binits{S.}},
\bauthor{\bsnm{Guarín-Zapata}, \binits{N.}}:
\batitle{Benchmarking physics-informed neural networks and {Boundary} {Element} {Method}: {Accuracy}–efficiency trade-offs in wave scattering}.
\bjtitle{Engineering Analysis with Boundary Elements}
\bvolume{188},
\bfpage{106786}
(\byear{2026})
\doiurl{10.1016/j.enganabound.2026.106786}
\end{barticle}
\endbibitem

\bibitem[\protect\citeauthoryear{Roncoroni et~al.}{2021}]{roncoroni_synthetic_2021}
\begin{botherref}
\oauthor{\bsnm{Roncoroni}, \binits{G.}},
\oauthor{\bsnm{Fortini}, \binits{C.}},
\oauthor{\bsnm{Bortolussi}, \binits{L.}},
\oauthor{\bsnm{Bienati}, \binits{N.}},
\oauthor{\bsnm{Pipan}, \binits{M.}}:
Synthetic seismic data generation with deep learning.
Journal of Applied Geophysics
\textbf{190}
(2021)
\doiurl{10.1016/j.jappgeo.2021.104347}
\end{botherref}
\endbibitem

\bibitem[\protect\citeauthoryear{Rincón et~al.}{2026a}]{rincon_2026_21054377}
\begin{botherref}
\oauthor{\bsnm{Rincón}, \binits{O.A.}},
\oauthor{\bsnm{Perez~Bernal}, \binits{G.}},
\oauthor{\bsnm{Montoya-Noguera}, \binits{S.}},
\oauthor{\bsnm{Guarín-Zapata}, \binits{N.}}:
Extracted Data from a Scoping Review of Machine Learning Approaches for Wave Propagation Modeling in Seismology.
\doiurl{10.5281/zenodo.21054377}
\end{botherref}
\endbibitem

\bibitem[\protect\citeauthoryear{Rincón et~al.}{2026b}]{rincon_2026_21017942}
\begin{botherref}
\oauthor{\bsnm{Rincón}, \binits{O.A.}},
\oauthor{\bsnm{Perez~Bernal}, \binits{G.}},
\oauthor{\bsnm{Montoya-Noguera}, \binits{S.}},
\oauthor{\bsnm{Guarín-Zapata}, \binits{N.}}:
Literature Dataset of Identified Studies for a Scoping Review of Physics-Informed Machine Learning for Wave Propagation Modeling in Seismology.
\doiurl{10.5281/zenodo.21017942}
\end{botherref}
\endbibitem

\bibitem[\protect\citeauthoryear{Rincón et~al.}{2026c}]{rincon_2026_20834562}
\begin{botherref}
\oauthor{\bsnm{Rincón}, \binits{O.A.}},
\oauthor{\bsnm{Perez~Bernal}, \binits{G.}},
\oauthor{\bsnm{Montoya-Noguera}, \binits{S.}},
\oauthor{\bsnm{Guarín-Zapata}, \binits{N.}}:
Literature Dataset of Selected for a Scoping Review of Physics-Informed Machine Learning for Wave Propagation Modeling in Seismology.
Zenodo
(2026).
\doiurl{10.5281/zenodo.20834562}
\end{botherref}
\endbibitem

\bibitem[\protect\citeauthoryear{Raissi et~al.}{2019}]{raissi_physics-informed_2019}
\begin{barticle}
\bauthor{\bsnm{Raissi}, \binits{M.}},
\bauthor{\bsnm{Perdikaris}, \binits{P.}},
\bauthor{\bsnm{Karniadakis}, \binits{G.E.}}:
\batitle{Physics-informed neural networks: {A} deep learning framework for solving forward and inverse problems involving nonlinear partial differential equations}.
\bjtitle{Journal of Computational Physics}
\bvolume{378},
\bfpage{686}--\blpage{707}
(\byear{2019})
\doiurl{10.1016/j.jcp.2018.10.045}
\end{barticle}
\endbibitem

\bibitem[\protect\citeauthoryear{Röth and Tarantola}{1994}]{roth_neural_1994}
\begin{botherref}
\oauthor{\bsnm{Röth}, \binits{G.}},
\oauthor{\bsnm{Tarantola}, \binits{A.}}:
Neural networks and inversion of seismic data.
Journal of Geophysical Research: Solid Earth
\textbf{99}(B4)
(1994)
\doiurl{10.1029/93JB01563}
\end{botherref}
\endbibitem

\bibitem[\protect\citeauthoryear{Rüde et~al.}{2018}]{rude_research_2018}
\begin{barticle}
\bauthor{\bsnm{Rüde}, \binits{U.}},
\bauthor{\bsnm{Willcox}, \binits{K.}},
\bauthor{\bsnm{McInnes}, \binits{L.C.}},
\bauthor{\bsnm{Sterck}, \binits{H.D.}}:
\batitle{Research and {Education} in {Computational} {Science} and {Engineering}}.
\bjtitle{SIAM Review}
\bvolume{60}(\bissue{3}),
\bfpage{707}--\blpage{754}
(\byear{2018})
\doiurl{10.1137/16M1096840}
\end{barticle}
\endbibitem

\bibitem[\protect\citeauthoryear{Ren et~al.}{2020}]{ren_physics-based_2020}
\begin{barticle}
\bauthor{\bsnm{Ren}, \binits{Y.}},
\bauthor{\bsnm{Xu}, \binits{X.}},
\bauthor{\bsnm{Yang}, \binits{S.}},
\bauthor{\bsnm{Nie}, \binits{L.}},
\bauthor{\bsnm{Chen}, \binits{Y.}}:
\batitle{A {Physics}-{Based} {Neural}-{Network} {Way} to {Perform} {Seismic} {Full} {Waveform} {Inversion}}.
\bjtitle{IEEE Access}
\bvolume{8},
\bfpage{112266}--\blpage{112277}
(\byear{2020})
\doiurl{10.1109/ACCESS.2020.2997921}
\end{barticle}
\endbibitem

\bibitem[\protect\citeauthoryear{Raissi et~al.}{2020}]{raissi_hidden_2020}
\begin{barticle}
\bauthor{\bsnm{Raissi}, \binits{M.}},
\bauthor{\bsnm{Yazdani}, \binits{A.}},
\bauthor{\bsnm{Karniadakis}, \binits{G.E.}}:
\batitle{Hidden fluid mechanics: {Learning} velocity and pressure fields from flow visualizations}.
\bjtitle{Science}
\bvolume{367}(\bissue{6481}),
\bfpage{1026}--\blpage{1030}
(\byear{2020})
\doiurl{10.1126/science.aaw4741}
\end{barticle}
\endbibitem

\bibitem[\protect\citeauthoryear{Seriani and Oliveira}{2020}]{seriani_numerical_2020}
\begin{barticle}
\bauthor{\bsnm{Seriani}, \binits{G.}},
\bauthor{\bsnm{Oliveira}, \binits{S.P.}}:
\batitle{Numerical modeling of mechanical wave propagation}.
\bjtitle{Rivista del Nuovo Cimento}
\bvolume{43}(\bissue{9}),
\bfpage{459}--\blpage{514}
(\byear{2020})
\doiurl{10.1007/s40766-020-00009-0}
\end{barticle}
\endbibitem

\bibitem[\protect\citeauthoryear{Stein and Wysession}{2009}]{stein_introduction_2009}
\begin{bbook}
\bauthor{\bsnm{Stein}, \binits{S.}},
\bauthor{\bsnm{Wysession}, \binits{M.}}:
\bbtitle{An {Introduction} to {Seismology}, {Earthquakes}, and {Earth} {Structure}}.
\bpublisher{Wiley-Blackwell},
\blocation{Malden, Mass. Berlin}
(\byear{2009})
\end{bbook}
\endbibitem

\bibitem[\protect\citeauthoryear{Song and Wang}{2022}]{song_high-frequency_2022}
\begin{barticle}
\bauthor{\bsnm{Song}, \binits{C.}},
\bauthor{\bsnm{Wang}, \binits{Y.}}:
\batitle{High-frequency wavefield extrapolation using the {Fourier} neural operator}.
\bjtitle{Journal of Geophysics and Engineering}
\bvolume{19}(\bissue{2}),
\bfpage{269}--\blpage{282}
(\byear{2022})
\doiurl{10.1093/jge/gxac016}
\end{barticle}
\endbibitem

\bibitem[\protect\citeauthoryear{Song and Wang}{2023}]{song_simulating_2023}
\begin{barticle}
\bauthor{\bsnm{Song}, \binits{C.}},
\bauthor{\bsnm{Wang}, \binits{Y.}}:
\batitle{Simulating seismic multifrequency wavefields with the {Fourier} feature physics-informed neural network}.
\bjtitle{Geophysical Journal International}
\bvolume{232}(\bissue{3}),
\bfpage{1503}--\blpage{1514}
(\byear{2023})
\doiurl{10.1093/gji/ggac399}
\end{barticle}
\endbibitem

\bibitem[\protect\citeauthoryear{Tarantola}{2005}]{tarantola2005inverse}
\begin{bbook}
\bauthor{\bsnm{Tarantola}, \binits{A.}}:
\bbtitle{Inverse Problem Theory and Methods for Model Parameter Estimation}.
\bpublisher{SIAM},
\blocation{Philadelphia}
(\byear{2005})
\end{bbook}
\endbibitem

\bibitem[\protect\citeauthoryear{Tricco et~al.}{2018}]{tricco_prisma_2018}
\begin{barticle}
\bauthor{\bsnm{Tricco}, \binits{A.C.}},
\bauthor{\bsnm{Lillie}, \binits{E.}},
\bauthor{\bsnm{Zarin}, \binits{W.}},
\bauthor{\bsnm{O'Brien}, \binits{K.K.}},
\bauthor{\bsnm{Colquhoun}, \binits{H.}},
\bauthor{\bsnm{Levac}, \binits{D.}},
\bauthor{\bsnm{Moher}, \binits{D.}},
\bauthor{\bsnm{Peters}, \binits{M.D.J.}},
\bauthor{\bsnm{Horsley}, \binits{T.}},
\bauthor{\bsnm{Weeks}, \binits{L.}},
\bauthor{\bsnm{Hempel}, \binits{S.}},
\bauthor{\bsnm{Akl}, \binits{E.A.}},
\bauthor{\bsnm{Chang}, \binits{C.}},
\bauthor{\bsnm{McGowan}, \binits{J.}},
\bauthor{\bsnm{Stewart}, \binits{L.}},
\bauthor{\bsnm{Hartling}, \binits{L.}},
\bauthor{\bsnm{Aldcroft}, \binits{A.}},
\bauthor{\bsnm{Wilson}, \binits{M.G.}},
\bauthor{\bsnm{Garritty}, \binits{C.}},
\bauthor{\bsnm{Lewin}, \binits{S.}},
\bauthor{\bsnm{Godfrey}, \binits{C.M.}},
\bauthor{\bsnm{Macdonald}, \binits{M.T.}},
\bauthor{\bsnm{Langlois}, \binits{E.V.}},
\bauthor{\bsnm{Soares-Weiser}, \binits{K.}},
\bauthor{\bsnm{Moriarty}, \binits{J.}},
\bauthor{\bsnm{Clifford}, \binits{T.}},
\bauthor{\bsnm{Tuncalp}, \binits{O.}},
\bauthor{\bsnm{Straus}, \binits{S.E.}}:
\batitle{{PRISMA} {Extension} for {Scoping} {Reviews} ({PRISMA}-{ScR}): {Checklist} and {Explanation}}.
\bjtitle{Annals of Internal Medicine}
\bvolume{169}(\bissue{7}),
\bfpage{467}--\blpage{473}
(\byear{2018})
\doiurl{10.7326/M18-0850}
\end{barticle}
\endbibitem

\bibitem[\protect\citeauthoryear{Vadyala et~al.}{2022}]{vadyala_review_2022}
\begin{barticle}
\bauthor{\bsnm{Vadyala}, \binits{S.R.}},
\bauthor{\bsnm{Betgeri}, \binits{S.N.}},
\bauthor{\bsnm{Matthews}, \binits{J.C.}},
\bauthor{\bsnm{Matthews}, \binits{E.}}:
\batitle{A review of physics-based machine learning in civil engineering}.
\bjtitle{Results in Engineering}
\bvolume{13},
\bfpage{100316}
(\byear{2022})
\doiurl{10.1016/j.rineng.2021.100316}
\end{barticle}
\endbibitem

\bibitem[\protect\citeauthoryear{Virieux et~al.}{2011}]{virieux_review_2011}
\begin{barticle}
\bauthor{\bsnm{Virieux}, \binits{J.}},
\bauthor{\bsnm{Calandra}, \binits{H.}},
\bauthor{\bsnm{Plessix}, \binits{R.-E.}}:
\batitle{A review of the spectral, pseudo-spectral, finite-difference and finite-element modelling techniques for geophysical imaging}.
\bjtitle{Geophysical Prospecting}
\bvolume{59}(\bissue{5}),
\bfpage{794}--\blpage{813}
(\byear{2011})
\doiurl{10.1111/j.1365-2478.2011.00967.x}
\end{barticle}
\endbibitem

\bibitem[\protect\citeauthoryear{Wang et~al.}{2023}]{wang_memory_2023}
\begin{botherref}
\oauthor{\bsnm{Wang}, \binits{S.}},
\oauthor{\bsnm{Jiang}, \binits{Y.}},
\oauthor{\bsnm{Song}, \binits{P.}},
\oauthor{\bsnm{Tan}, \binits{J.}},
\oauthor{\bsnm{Liu}, \binits{Z.}},
\oauthor{\bsnm{He}, \binits{B.}}:
Memory {Optimization} in {RNN}-{Based} {Full} {Waveform} {Inversion} {Using} {Boundary} {Saving} {Wavefield} {Reconstruction}.
IEEE Transactions on Geoscience and Remote Sensing
\textbf{61}
(2023)
\doiurl{10.1109/TGRS.2023.3317529}
\end{botherref}
\endbibitem

\bibitem[\protect\citeauthoryear{Wang et~al.}{2018}]{wang_velocity_2018}
\begin{bchapter}
\bauthor{\bsnm{Wang}, \binits{W.}},
\bauthor{\bsnm{Yang}, \binits{F.}},
\bauthor{\bsnm{Ma}, \binits{J.}}:
\bctitle{Velocity model building with a modified fully convolutional network},
pp. \bfpage{2086}--\blpage{2090}
(\byear{2018}).
\doiurl{10.1190/segam2018-2997566.1}
\end{bchapter}
\endbibitem

\bibitem[\protect\citeauthoryear{Xu et~al.}{2019}]{xu_physics_2019}
\begin{bchapter}
\bauthor{\bsnm{Xu}, \binits{Y.}},
\bauthor{\bsnm{Li}, \binits{J.}},
\bauthor{\bsnm{Chen}, \binits{X.}}:
\bctitle{Physics informed neural networks for velocity inversion},
pp. \bfpage{2584}--\blpage{2588}
(\byear{2019}).
\doiurl{10.1190/segam2019-3216823.1}
\end{bchapter}
\endbibitem

\bibitem[\protect\citeauthoryear{Xiong et~al.}{2022}]{xiong_deep-neural-networks-based_2022}
\begin{barticle}
\bauthor{\bsnm{Xiong}, \binits{F.}},
\bauthor{\bsnm{Liu}, \binits{J.}},
\bauthor{\bsnm{Guo}, \binits{Z.}},
\bauthor{\bsnm{Liu}, \binits{J.}}:
\batitle{Deep-neural-networks-based approaches for {Biot}–squirt model in rock physics}.
\bjtitle{Acta Geophysica}
\bvolume{70}(\bissue{2}),
\bfpage{593}--\blpage{607}
(\byear{2022})
\doiurl{10.1007/s11600-022-00740-8}
\end{barticle}
\endbibitem

\bibitem[\protect\citeauthoryear{Yang et~al.}{2023}]{yang_rapid_2023}
\begin{botherref}
\oauthor{\bsnm{Yang}, \binits{Y.}},
\oauthor{\bsnm{Gao}, \binits{A.F.}},
\oauthor{\bsnm{Azizzadenesheli}, \binits{K.}},
\oauthor{\bsnm{Clayton}, \binits{R.W.}},
\oauthor{\bsnm{Ross}, \binits{Z.E.}}:
Rapid {Seismic} {Waveform} {Modeling} and {Inversion} {With} {Neural} {Operators}.
IEEE Transactions on Geoscience and Remote Sensing
\textbf{61}
(2023)
\doiurl{10.1109/TGRS.2023.3264210}
\end{botherref}
\endbibitem

\bibitem[\protect\citeauthoryear{Yang et~al.}{2021}]{yang_seismic_2021}
\begin{barticle}
\bauthor{\bsnm{Yang}, \binits{Y.}},
\bauthor{\bsnm{Gao}, \binits{A.F.}},
\bauthor{\bsnm{Castellanos}, \binits{J.C.}},
\bauthor{\bsnm{Ross}, \binits{Z.E.}},
\bauthor{\bsnm{Azizzadenesheli}, \binits{K.}},
\bauthor{\bsnm{Clayton}, \binits{R.W.}}:
\batitle{Seismic {Wave} {Propagation} and {Inversion} with {Neural} {Operators}}.
\bjtitle{Seismic Record}
\bvolume{1}(\bissue{3}),
\bfpage{126}--\blpage{134}
(\byear{2021})
\doiurl{10.1785/0320210026}
\end{barticle}
\endbibitem

\bibitem[\protect\citeauthoryear{Yang and Ma}{2021}]{yang_revisit_2021}
\begin{botherref}
\oauthor{\bsnm{Yang}, \binits{F.}},
\oauthor{\bsnm{Ma}, \binits{J.}}:
Revisit {Geophysical} {Imaging} in {A} {New} {View} of {Physics}-informed {Generative} {Adversarial} {Learning}.
arXiv
(2021).
\doiurl{10.48550/arXiv.2109.11452}
\end{botherref}
\endbibitem

\bibitem[\protect\citeauthoryear{Zou et~al.}{2023}]{zou_deep_2023}
\begin{botherref}
\oauthor{\bsnm{Zou}, \binits{C.}},
\oauthor{\bsnm{Azizzadenesheli}, \binits{K.}},
\oauthor{\bsnm{Ross}, \binits{Z.E.}},
\oauthor{\bsnm{Clayton}, \binits{R.W.}}:
Deep {Neural} {Helmholtz} {Operators} for {3D} {Elastic} {Wave} {Propagation} and {Inversion}.
arXiv
(2023).
\doiurl{10.48550/arXiv.2311.09608}
\end{botherref}
\endbibitem

\bibitem[\protect\citeauthoryear{Zhang et~al.}{2023}]{zhang_autoencoded_2023}
\begin{barticle}
\bauthor{\bsnm{Zhang}, \binits{C.}},
\bauthor{\bsnm{Li}, \binits{J.}},
\bauthor{\bsnm{Yu}, \binits{H.}},
\bauthor{\bsnm{Liu}, \binits{B.}}:
\batitle{Autoencoded {Elastic} {Wave}-{Equation} {Traveltime} {Inversion}: {Toward} {Reliable} {Near}-{Surface} {Tomogram}}.
\bjtitle{IEEE Transactions on Geoscience and Remote Sensing}
\bvolume{61},
\bfpage{1}--\blpage{13}
(\byear{2023})
\doiurl{10.1109/TGRS.2023.3243140}
\end{barticle}
\endbibitem

\bibitem[\protect\citeauthoryear{Zou et~al.}{2024}]{zou_seismic_2024}
\begin{botherref}
\oauthor{\bsnm{Zou}, \binits{J.}},
\oauthor{\bsnm{Liu}, \binits{C.}},
\oauthor{\bsnm{Zhao}, \binits{P.}},
\oauthor{\bsnm{Song}, \binits{C.}}:
Seismic {Wavefields} {Modeling} {With} {Variable} {Horizontally} {Layered} {Velocity} {Models} via {Velocity}-{Encoded} {PINN}.
IEEE Transactions on Geoscience and Remote Sensing
\textbf{62}
(2024)
\doiurl{10.1109/TGRS.2024.3411472}
\end{botherref}
\endbibitem

\bibitem[\protect\citeauthoryear{Zubov et~al.}{2021}]{zubov_neuralpde_2021}
\begin{botherref}
\oauthor{\bsnm{Zubov}, \binits{K.}},
\oauthor{\bsnm{McCarthy}, \binits{Z.}},
\oauthor{\bsnm{Ma}, \binits{Y.}},
\oauthor{\bsnm{Calisto}, \binits{F.}},
\oauthor{\bsnm{Pagliarino}, \binits{V.}},
\oauthor{\bsnm{Azeglio}, \binits{S.}},
\oauthor{\bsnm{Bottero}, \binits{L.}},
\oauthor{\bsnm{Luján}, \binits{E.}},
\oauthor{\bsnm{Sulzer}, \binits{V.}},
\oauthor{\bsnm{Bharambe}, \binits{A.}},
\oauthor{\bsnm{Vinchhi}, \binits{N.}},
\oauthor{\bsnm{Balakrishnan}, \binits{K.}},
\oauthor{\bsnm{Upadhyay}, \binits{D.}},
\oauthor{\bsnm{Rackauckas}, \binits{C.}}:
{NeuralPDE}: {Automating} {Physics}-{Informed} {Neural} {Networks} ({PINNs}) with {Error} {Approximations}.
arXiv.
arXiv:2107.09443 [cs.MS]
(2021).
\doiurl{10.48550/arXiv.2107.09443}
\end{botherref}
\endbibitem

\bibitem[\protect\citeauthoryear{Zhang et~al.}{2024}]{zhang_seismic_2024}
\begin{barticle}
\bauthor{\bsnm{Zhang}, \binits{Y.}},
\bauthor{\bsnm{Meng}, \binits{D.}},
\bauthor{\bsnm{Zhou}, \binits{Y.}},
\bauthor{\bsnm{Song}, \binits{L.}},
\bauthor{\bsnm{Dong}, \binits{H.}}:
\batitle{Seismic {Velocity} {Inversion} {Based} on {Physically} {Constrained} {Neural} {Networks}}.
\bjtitle{IEEE Transactions on Geoscience and Remote Sensing}
\bvolume{62},
\bfpage{1}--\blpage{17}
(\byear{2024})
\doiurl{10.1109/TGRS.2023.3339783}
\end{barticle}
\endbibitem

\bibitem[\protect\citeauthoryear{Zhang et~al.}{2023}]{zhang_physics-guided_2023}
\begin{barticle}
\bauthor{\bsnm{Zhang}, \binits{Y.}},
\bauthor{\bsnm{Singh}, \binits{S.}},
\bauthor{\bsnm{Thanoon}, \binits{D.}},
\bauthor{\bsnm{Devarakota}, \binits{P.}},
\bauthor{\bsnm{Jin}, \binits{L.}},
\bauthor{\bsnm{Tsvankin}, \binits{I.}}:
\batitle{Physics-{Guided} {Unsupervised} {Deep}-{Learning} {Seismic} {Inversion} {With} {Uncertainty} {Quantification}}.
\bjtitle{Journal of Seismic Exploration}
\bvolume{32}(\bissue{3}),
\bfpage{257}--\blpage{270}
(\byear{2023})
\end{barticle}
\endbibitem

\bibitem[\protect\citeauthoryear{Zhang et~al.}{2020}]{zhang_survey_2020}
\begin{botherref}
\oauthor{\bsnm{Zhang}, \binits{Y.}},
\oauthor{\bsnm{Tiňo}, \binits{P.}},
\oauthor{\bsnm{Leonardis}, \binits{A.}},
\oauthor{\bsnm{Tang}, \binits{K.}}:
A {Survey} on {Neural} {Network} {Interpretability}
(2020).
\doiurl{10.1109/TETCI.2021.3100641}
\end{botherref}
\endbibitem

\bibitem[\protect\citeauthoryear{Zhang et~al.}{2023}]{zhang_seismic_2023}
\begin{barticle}
\bauthor{\bsnm{Zhang}, \binits{Y.}},
\bauthor{\bsnm{Zhu}, \binits{X.}},
\bauthor{\bsnm{Gao}, \binits{J.}}:
\batitle{Seismic {Inversion} {Based} on {Acoustic} {Wave} {Equations} {Using} {Physics}-{Informed} {Neural} {Network}}.
\bjtitle{IEEE Transactions on Geoscience and Remote Sensing}
\bvolume{61},
\bfpage{1}--\blpage{11}
(\byear{2023})
\doiurl{10.1109/TGRS.2023.3236973}
\end{barticle}
\endbibitem

\end{thebibliography}

\end{document}